\documentclass[fleqn,10pt]{wlscirep}
\usepackage[utf8]{inputenc}
\usepackage[T1]{fontenc}
\usepackage{mathtools}
\usepackage{siunitx}
\DeclarePairedDelimiterXPP\BigOSI[2]%
  {\mathcal{O}}{(}{)}{}%
  {\SI{#1}{#2}}

\usepackage{tabularx}
\usepackage{comment}
\usepackage{color}
\usepackage{chemformula}

\graphicspath{ {img/} }

\title{Potential of laser-driven VHEEs toward FLASH radiotherapy: Monte Carlo dosimetric study of single-field pencil beam scanning of a brain tumor}

\author[1,2,*]{Leonida A. Gizzi}
\author[3]{Damiano Del Sarto}
\author[1]{Federico Avella}
\author[1]{Gabriele Bandini}
\author[1]{Simona Piccinini}
\author[4]{Daniele Panetta}
\author[5]{Davide Terzani}
\author[1,2,**]{Luca Labate}

\affil[1]{Intense Laser Irradiation Laboratory, CNR-INO, Pisa, Italy}
\affil[2]{Istituto Nazionale di Fisica Nucleare, Sezione di Pisa, Italy}
\affil[3]{now at Clinica S.Rossore, Pisa, Italy}
\affil[4]{CNR-IFC, Pisa, Italy}
\affil[5]{Lawrence Berkeley National Laboratory, Berkeley, USA}
\affil[*]{la.gizzi@ino.cnr.it}
\affil[**]{luca.labate@ino.cnr.it}

\begin{abstract}
Radiotherapy with Very High Energy Electron (VHEE) beams is being extensively investigated for the treatment of deep-seated tumours, even in view of novel protocols based on the so-called FLASH effect.
Laser WakeField Acceleration (LWFA) provides a compact and affordable accelerator technology for VHEE electron beams, featuring ultra-high instantaneous dose rates and holding the promise to provide Ultra-High (average) Dose Rates (UHDRs) needed to activate the FLASH effect, with major efforts ongoing worldwide to fulfill this promise.
Therapeutic doses are already at reach, using pencil beams produced via LWFA. These beams typically exhibit significant energy spread, and small transverse size.
These features are rather different from those of other  beams considered so far in radiotherapy studies.
In view of a rapid clinical translation of LWFA-VHEE beams it is therefore of paramount importance to assess the role of these properties in the dose delivery to the patient.
Here we present a study carried out via start-to-end (PIC and Monte Carlo) simulations, of the main dosimetric features of a realistic laser-driven VHEE pencil beam targeted on a brain tumor.
The entire tumor coverage is achieved by a scanning procedure; the dose pattern resulting from tessellation, i.e. the overlapping of adjacent beamlets, and the role of energy spread are thoroughly discussed.
Dose Volume Histograms are presented, and their quality is discussed.
The impact of the FLASH effect is also considered, introducing a degree of healthy tissue sparing in the modelling.
Finally, the foreseen technological path toward the achievement of FLASH dose rates with LWFA-VHEE beams is briefly outlined.
\end{abstract} 
\begin{document}

\flushbottom
\maketitle
%
%
\thispagestyle{empty}


\section*{Introduction and motivations}

Radiotherapy (RT) is an essential tool for cancer treatment, with more than 60\% of new cancer patients in high income countries expected to undergo a RT treatment as part of their cancer management \cite{zhu2024global}.
RT relies on the possibility of inducing lethal damage to tumor cells while keeping the adverse effects probability to the surrounding healthy tissue at an acceptable level.
This is made possible by the usage of advanced dose conformation techniques and relies, ultimately, on the DNA repairing mechanisms operating more efficiently in healthy tissues rather than in tumors \cite{Chen2022}.
The so-called therapeutic window identifies the useful range of irradiation dose for which the probability of tumor control exceeds the probability of normal tissue damage, once repairing mechanisms take place.
The quest for mechanisms able to increase this window is an active research field, whose findings would enable better treatments, in particular in the case of radio-resistant tumors and/or of tumors localized near sensitive organs.
In this respect, the so-called FLASH effect \cite{Favaudon_2014} has sparked hope for a possible breakthrough in RT (for reviews on the matter see \cite{Panaino_2024_cancers17020181, Vozenin2019, RevModPhys.96.035002}).
It consists in a significant reduction of damage to healthy tissue at a given dose when ionizing radiation is delivered at ultra high dose rates (UHDR), of the order of several tens of $\mathrm{Gy}/\mathrm{s}$ or more, orders of magnitude higher than the rates ($\sim \mathrm{Gy}/\mathrm{min}$) used in current RT protocols.
Although the fundamental mechanisms underpinning the FLASH effect are still to be clarified \cite{Vozenin_2025_ps, Kacem_IJRB2021, Friedl_MP2022}, a sparing of healthy tissue at ultra-high dose rates with preserved tumor control has now been confirmed in several studies (see \cite{mcgarrigle2024flash, Borghini_2022, Borghini2024_FLASH, Chow2024_FLASH_Mechanisms, Rosini2025_FLASH_Mechanisms, Wilson_FO2020a} and Refs. therein).\par

Crucial for the clinical translation of FLASH-based RT protocols are the properties of the radiation sources.
Indeed, the dose rate needed for FLASH treatments requires high beam current, which is out of the range of operation of clinical accelerators used for the treatment of deep seated tumors, mainly because of the poor intrinsic conversion efficiency of the underlying \textit{Bremsstrahlung} process leading to the generation of the deep penetrating photon beam starting from the multi MeV primary electron beam.
Experiments carried out so far to investigate the FLASH effect, either \textit{in vitro}, \textit{in vivo} or as a compassionate treatment of superficial tumors in humans, have been carried out using conventional devices modified to provide ultra-high dose rates (see \cite{Sampayan_SR2021, Esplen_PMB2020} and Refs. therein).
In the vast majority of cases, electron beams with relatively low energy have been used, only suitable for irradiation at shallow depths \cite{Wilson_FO2020a}, such as in the case of skin \cite{bourhis2019treatment}, eyes, salivary glands tumors \cite{Ronga_Cancers2021}, or in IORT \cite{Righi_JACMP2013}.
Electron beams capable of delivering FLASH dose rates, for instance, have been obtained by standard medical accelerators, by removing the electron-to-photon conversion element and tuning the beam forming and accelerating structures to sustain the needed high currents, and to provide, at the same time, beams with a suitable quality for radiobiology and medical experiments \cite{Lempart_RaO2019,Rahman_IJoRO2021,Felici_FP2020}; electrons with a maximum energy up to $\sim20\,\mathrm{MeV}$ can be obtained in this way.
Moreover, electron accelerators used for Intra Operative Radiation Therapy (IORT) are also available with FLASH dose-rates, with energy up to $\sim10\,\mathrm{MeV}$ and cm-wide beam size. 
These and other similar devices provide electron beams with UHDR with sufficient energy to irradiate superficial tumors and should be available soon for clinical applications.
In contrast, the treatment of deep seated tumors at FLASH dose-rates would require much more penetrating beams. 
Availability of medical accelerators of charged particle beams with both high energy (high penetration depth) and current (UHDR) is hindered by size, complexity and cost issues, although efforts for technology development are ongoing.
For instance, selected hadron therapy centers are currently capable of delivering FLASH dose-rate pencil beams and are starting to pave the way towards human clinical applications.

In this context, the use of another type of particle beam, namely Very High Energy Electrons (VHEE), with energy in the range between $100$ and $250\,\mathrm{MeV}$, is considered extremely promising for clinical applications.
The development of VHEE devices based on RF technology is currently actively pursued by different companies, usually in a joint effort with government laboratories \cite{THERIQ, Lumitron}.
An updated discussion of the ongoing research activity in this field can be found in \cite{Farr_MP2022} and in a very recent review \cite{Panaino_2024_cancers17020181}.
As a matter of fact, originally, the use of high energy electrons to irradiate deep seated tumors was mostly investigated via theoretical/numerical studies.
Up until the beginning of 2000s, the energy range considered was limited to $\sim 50\,\mathrm{MeV}$.
More recently, in light of both technology advancements in conventional acceleration \cite{
DesRosiers_PMB2000,Yeboah_PMB2002,Palma_RaO2016} and the emergence of laser-driven electron acceleration, dosimetric studies of VHEE beams started to be performed (see for instance \cite{Ronga_Cancers2021} and Refs. therein).
Monte Carlo simulations, in particular, have shown appealing features of VHEE, such as their reduced lateral spread in tissues with respect to both $\lesssim 50\,\mathrm{MeV}$ e-beams and high energy protons (also in the case of FLASH modality \cite{Muscato_2023, Bohlen2024VHEEFLASH, Gesualdi2025ProtonVHEE}); furthermore, the quality of the achievable dose conformation for some types of tumors has been shown (numerically) to be comparable or even superior to that of clinical VMAT protocols\cite{BazalovaCarter_MP2015,Fuchs_PMB2009,Rahman_RaO2022}.
Moreover, the remarkable resilience of VHEE beams to tissue inhomogeneities has been pointed out and showed to be an added value with respect to photon and proton beams \cite{Lagzda_NIMB2020,DesRosiers_SPIE2008}.
Other useful features offered by VHEE beams have also been recently highlighted, including the  possibility of fast magnetic steering to follow the patient physiological motion \cite{Maxim_RaO2019,Kokurewicz_SR2019}, or of magnetic focusing to locally enhance the delivered dose, thus providing an additional degree of freedom to optimize dose conformation\cite{Kokurewicz_SR2019}.

\subsection*{RT with laser-driven VHEE: basics and main issues}
In this paper we investigate the possibility of scanning a deep seated tumor volume using a VHEE pencil beam obtained via  Laser WakeField Acceleration (LWFA)\cite{Esarey_RMP2009}.
The LWFA process enables very high accelerating gradients to be established in a plasma, up to 3 orders of magnitude higher than in a typical RF LINAC, thus leading to a corresponding reduction in size of the accelerator stage.
This makes laser-driven accelerators potentially very compact ("table-top") and thus very attractive for a widespread clinical use.
LWFA acceleration of beams with energy from $\sim 100\,\mathrm{MeVs}$ up to $\sim8\,\mathrm{GeV}$  has been demonstrated experimentally in millimeter to few tens of centimeters long plasmas\cite{Gonsalves_PRL2019}.
In particular, the VHEE energy range is nowadays routinely accessed in many ultraintense laser laboratories hosting $100 \,\mathrm{TW}$ scale laser systems \cite{Danson_HPLSE2019,Gizzi_HPLSE2021}. Recently, high stability and reproducibility of such LWFA electron beams has been demostrated\cite{Maier_PRX2020}, making laser-driven accelerators sufficiently mature for application in RT.
Furthermore, LWFA beams are intrinsically characterized by ultrashort (sub-ps) duration, which translates into high intensity beams with instantaneous dose rates up to $10^{12}-10^{13}\,\mathrm{Gy/s}$.
Biological consequences of this feature, although  implications are still to be investigated in depth, had been already studied  in early radiobiology experiments with laser-driven beams \cite{Giulietti_2016} and more recently highlighted in novel studies \cite{Vozenin_2025_ps}.
This latter study with VHEE beams identified the role of beam intensity as a key parameter in the activation of the FLASH effect, showing that the higher the dose rate, the lower the dose needed to trigger the FLASH sparing effect.
These results support FLASH operation at a small dose per fraction in the clinical setting, where fractionation  remains a standard of care.
\par

\begin{figure}
    \begin{center}
    \includegraphics[width=0.5\columnwidth]{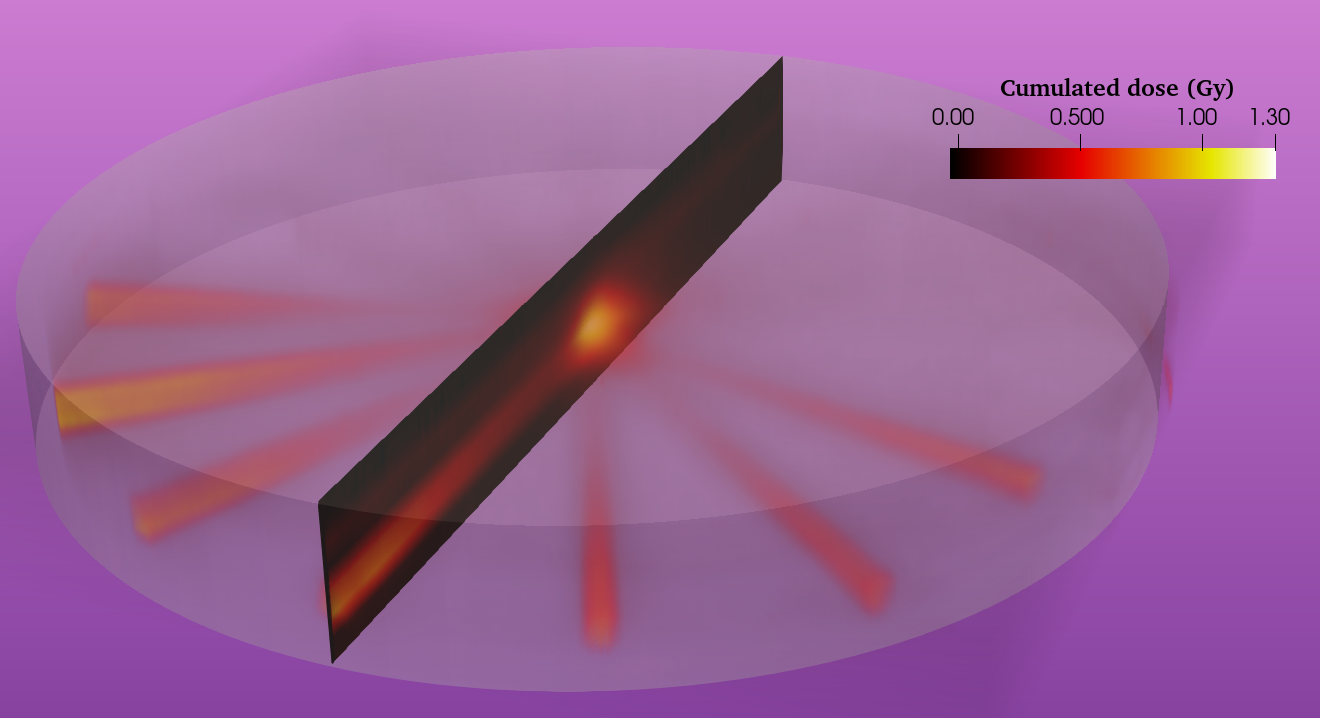}
    \end{center}
    \caption{3D dose pattern obtained in a recent experiment with laser-driven VHEE beams. The target volume was placed at the center of cylindrical phantom with 10cm diameter. Beam features were the same as those reported in \cite{Labate_SR2020}.}
    \label{fig::EXP_MF7}
\end{figure}

From an experimental viewpoint, efforts are ongoing to develop reliable compact, engineered laser-driven accelerators for VHEE RT\cite{Li_FP2025,Guo_NC2025}.
In parallel, dosimetric studies are being carried out.
For instance, the possibility of achieving dose conformation via advanced multi-field or intensity modulation techniques, typical of current RT protocols, has been shown in \cite{Labate_SR2020}, and the use of magnetic guiding based on permanent quadrupoles for pointing stabilization, transport and focusing to a phantom has been reported in \cite{Svendsen_SR2021}.
As an example of a recent experiment with laser-driven VHEE beams,  
Figure \ref{fig::EXP_MF7} shows the dose pattern obtained at the Intense Laser Irradiation Lab (ILIL) of CNR-INO in Pisa \cite{Gizzi_HPLSE2021},
when targeting a deep-seated, mm-size volume in the center of a 10cm size cylindrical phantom.\par

The results mentioned above, both theoretical and experimental, make a compelling case for the investigation of the dosimetric features of existing laser-driven VHEE beams.
With this respect, two aspects characterizing these beams, namely the small transverse size (few mm) and the quite large energy spread, deserve further studies, since they can strongly affect the  strategies for dose pattern optimization, making these strategies potentially different from well established techniques (and algorithms) employed in nowadays RT.
In principle, both the above peculiarities can be get rid of by a suitable beam conditioning. 
On the other hand, for the sake of a fast route toward the achievement of UHDR with laser-driven accelerators,
 using a pencil beam with few mm size and retaining all the charge  in the entire VHEE range, may represent an advantage.
It is thus worthwhile and timely to explore the dosimetry of beams with  such intrinsic features.
Here we report on start-to-end simulations of  a case study of irradiation of a brain tumor using a scanning technique, starting with a VHEE pencil beam as that used in \cite{Labate_SR2020}, which can be produced using existing, well validated LWFA acceleration regimes with 100 TW-class laser systems.
We observe that theoretica/numerical works carried out so far only reported on the use of "ideal" VHEE beams, with monochromatic energy and/or large transverse size \cite{BazalovaCarter_MP2015,Sarti_FO2021}; in contrast, our study considers beam features that have been demonstrated experimentally.

As a first step toward more advanced studies, we thus consider a 1-field (i.e., one direction) irradiation of a deep seated brain tumor.
While this may seem an oversimplified configuration compared to the state of the art of modern RT based on multi-field irradiation and/or intensity modulation, it is a much needed step and a building block toward more sophisticated treatment plannings.
To our knowledge no similar studies have been reported so far with such realistic VHEE pencil beams.
It is also worth pointing out that 
our study is not limited to the use of laser-driven VHEE accelerators and may be of more general relevance to VHEE beams accelerated using conventional devices.
Indeed, as mentioned, the requirement of a  UHDR to activate the FLASH effect naturally leads to considering higher bunch charge, thus relaxing the beam energy spread and/or to use beams with smaller sizes (pencil beams) than  the ones typical of current, photon based RT.\par

In the following Section, we first briefly motivate our choice of the tumor model for this case study, and the design of both the beam properties and the transport beamline.
Then we report on the main findings of our study, and finally, in light of the conclusions, we discuss perspectives for LWFA-based FLASH RT.

\section*{A case study: Tumor type and scanning pencil beam formation and conditioning}

A brain tumor is a particularly relevant case study to start with, in light of the major advancements that  a treatment exploiting the FLASH effect may potentially offer with respect to current RT protocols\cite{Mann_FN2018}.
Primary brain tumors occur in around 300000 people a year worldwide, accounting for 3.5\% of new cancer cases; in the pediatric age cohort (age 0-14) brain tumors are the second most common form of cancer \cite{globcan}.
Of all brain tumors, the glioblastoma (GBM) is the most common and aggressive one \cite{louis2016}, with a median survival of 15 months \cite{koshy2012}.
Current treatment protocols involve a multi-centered approach: surgical resection, radiation therapy and chemotherapy \cite{stupp2005}.
While the benefits of irradiating GBM are well established \cite{barani2015}, the radiation induced toxicity to the healthy tissues surrounding the lesions leads to neurological complications (e.g., learning and memory impairing, mood disorders, hearing loss, etc.); these complications are particularly severe for children \cite{denunzio2020, butler2006}.
It is well recognized \cite{montay2021} that the FLASH effect has the potential to significantly improve the risk-to-benefit ratio of RT for GBM.
In particular, the healthy tissue sparing effect of FLASH irradiation modality could prove essential to achieve better tumor control by increasing the delivered dose to the tumor while reducing the negative neurological side effects.
{\it In vivo} studies carried out so far on mice showed that FLASH RT is indeed an effective treatment against brain tumors \cite{montay2021, montay2017, alaghband2020}. \par

As a VHEE source, we consider a laser wakefield accelerator driven by a 100 TW-class ultrashort laser. 
It is worth stressing, at this point, that in our study we purposefully take into account laser (power, duration, energy per pulse) and VHEE beams (energy spread, divergence and/or emittance, etc.) parameters already available in laser laboratories worldwide, in order to keep the foreseen time horizon for the translation to clinical practice as short as possible.
On the other hand, as it comes out from this and other studies,  laser and LWFA technology developments are still required to access the UHDR regime needed for RT protocols exploiting the FLASH effect; these points will be discussed at the end of the paper.
We also mention that the production of electron beams such as those considered here have been routinely experimentally demonstrated at 100 TW class laser laboratories worldwide over the past decade.\par

Start-to-end simulations have been used in our study to make detailed beam statistics available for Monte Carlo simulations.
In particular, the LWFA process was simulated using a Particle-In-Cell (PIC) code, whereas Monte Carlo simulations were carried out to study the beam post-acceleration  manipulation, transport and interaction with a brain model.
The codes used for the two purposes are briefly presented in the Materials and Methods section (M\&M from now on).
In order to keep this paper as compact as possible, we refer to the Supplementary Materials (SM from now on) for a thorough discussion of the LWFA stage and of the accelerated beam properties.
Here we only mention that in our work we aimed at an acceleration regime providing a beam with most of the charge in the energy range $\sim 100-250\,\mathrm{MeV}$.
We observe that, although advanced LWFA acceleration methods able to deliver lower energy spread beams than those considered here have been developed over the past decade\cite{Pollock_PRL2011,Buck_PRL2013,Gonsalves_NP2011,Thaury_SR2015,Tomassini_PRAB2019}, a trade-off exists between the energy spread and the total charge per bunch (i.e., per laser shot), which determines the ultimate peak dose rate.
On the other hand, recent Monte Carlo simulations reported in the literature have shown a relatively weak dependence of the dose deposition pattern upon the energy spread of electrons with energy  $\gtrsim 100\,\mathrm{MeV}$; in light of this, our acceleration regime yields a charge as high as possible, allowing the beam spectrum to span the entire VHEE range.
The implications of this choice for future development of UHDR treatments, as well as the constraints on the total charge of electrons with energy lower than $\sim 100\,\mathrm{MeV}$, will be discussed later in the paper.\par

Table \ref{table:BeamParameters} (left column) summarizes the figures of the VHEE beam at the exit of the plasma (accelerator stage).
In particular, the bunch charge retrieved from the PIC simulations for electrons with energy $> 100 \,\mathrm{MeV}$ is of about $120\,\mathrm{pC}$ (the electron spectrum resulting from the PIC simulations is discussed in the SM).
It is worth mentioning here that a non-negligible lower energy component is in general to be expected from laser-driven accelerators.
Since this component is detrimental for RT, an energy selector to remove low energy electrons may be required.
Although a  discussion of such a selector  is  beyond the scope of this paper, we mention here that 
it may be obtained using 
a compact beamline, consisting of permanent magnets dipoles or other magnetic components.
\par

\begin{table}[]
    \centering
    {\footnotesize
    \begin{tabularx}{\textwidth}{|X|r||X|r||X|r|}
        \hline
        \textbf{Beam parameter} & \textbf{Value}  & \textbf{Beam parameter} & \textbf{Value} 
        & \textbf{Dose parameter} &
        \textbf{Value} \\
        \hline\hline
        Charge per bunch (laser shot) with $E\geq 100\,\mathrm{MeV}$ & $\sim 120\,\mathrm{pC}$ & Charge per bunch after the collimator & $\sim 50\,\mathrm{pC}$ &
        Trasverse size at buildup & $5.06\,\mathrm{mm}$\\
        \hline
        Divergence (RMS) & $3\,\mathrm{mrad}$ & Beam size at the collimator exit & $2.5\times2.5\,\mathrm{mm^2}$
        & Flatness at buildup & $<6.5\%$\\
        \hline
        Mean energy/energy spread & $\sim 220 \, \text{MeV}$ & Source-to-collimator distance & $380\,\mathrm{mm}$ 
        & Symmetry at buildup & $<2\%$ \\
        \hline
        & & Total source-to-patient distance & $700\,\mathrm{mm}$ 
        & &\\
        \hline
    \end{tabularx}
    }
    \caption{Table summarizing \textit{left column)} the VHEE beam parameters at LWFA stage exit, \textit{middle column}) the VHEE beam parameters downstream of the transport and collimator line, and \textit{right column}) the dose pattern figures at the phantom position.}
    \label{table:BeamParameters}
\end{table}

As reported in the SM, the VHEE beam has a nearly gaussian angular shape upon exiting the plasma.
In order to optimize the (transverse) dose pattern delivery, an \textit{ad hoc} collimator was designed by means of Monte Carlo simulations (see M\&M).
The rationale for the collimator design was the selection of a pencil beam with a sufficiently uniform transverse profile to enable a smooth coverage of the target volume, i.e., with minimal local dose overshoots in the overlapping regions between adjacent beams (see the next Section).
In addition, the structure  of the collimator was optimized so as to reduce the downstream dose contamination due to \textit{Bremsstrahlung} photons in the $X/\gamma$-ray region.
The final collimator structure is described in M\&M.
On exiting the collimator, the VHEE beam exhibits a square (flat-top) transverse profile, with the parameters shown in Table \ref{table:BeamParameters} (middle column).
The  beam is then transported in vacuum through a $300\,\mathrm{mm}$ long cylindrical pipe terminating with a kapton window (see M\&M).
This allowed a reasonable distance (total of approximately $700\,\mathrm{mm}$) to be maintained between the source and the patient, to be used for the beam steering in a real application, while ensuring that minimal perturbation to the shape of the beam occurs.
Finally, a further distance in air of $50\,\mathrm{mm}$ was considered up to the target (either water phantom or patient).
A sketch of the elements simulated in our Monte Carlo code downstream of the plasma "accelerator stage" is shown in Figure 2 \textit{left}.\par


\begin{figure}[!tbp]
  \centering
  \begin{minipage}[b]{0.55\textwidth}
    \includegraphics[width=\textwidth]{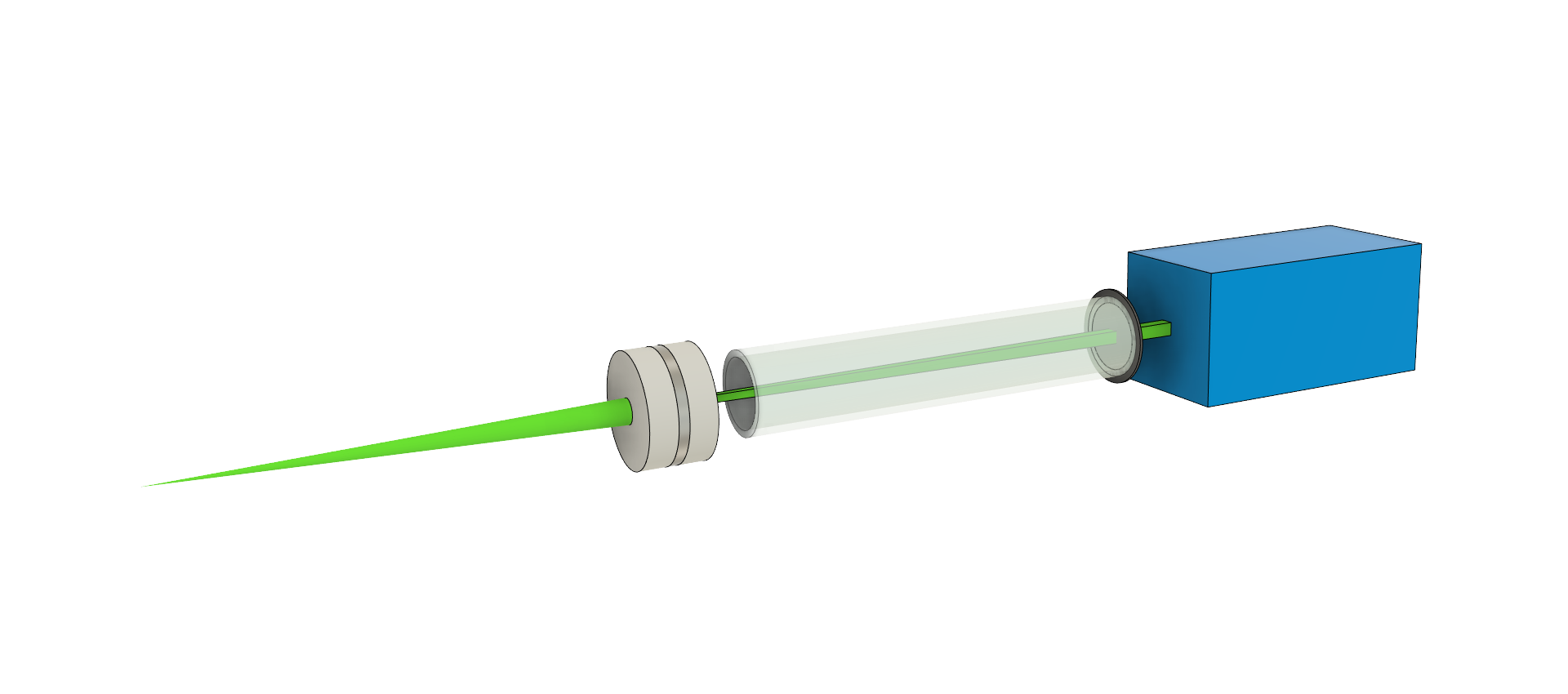}
  \end{minipage}
  \hfill
  \begin{minipage}[b]{0.4\textwidth}
    \includegraphics[width=\textwidth]{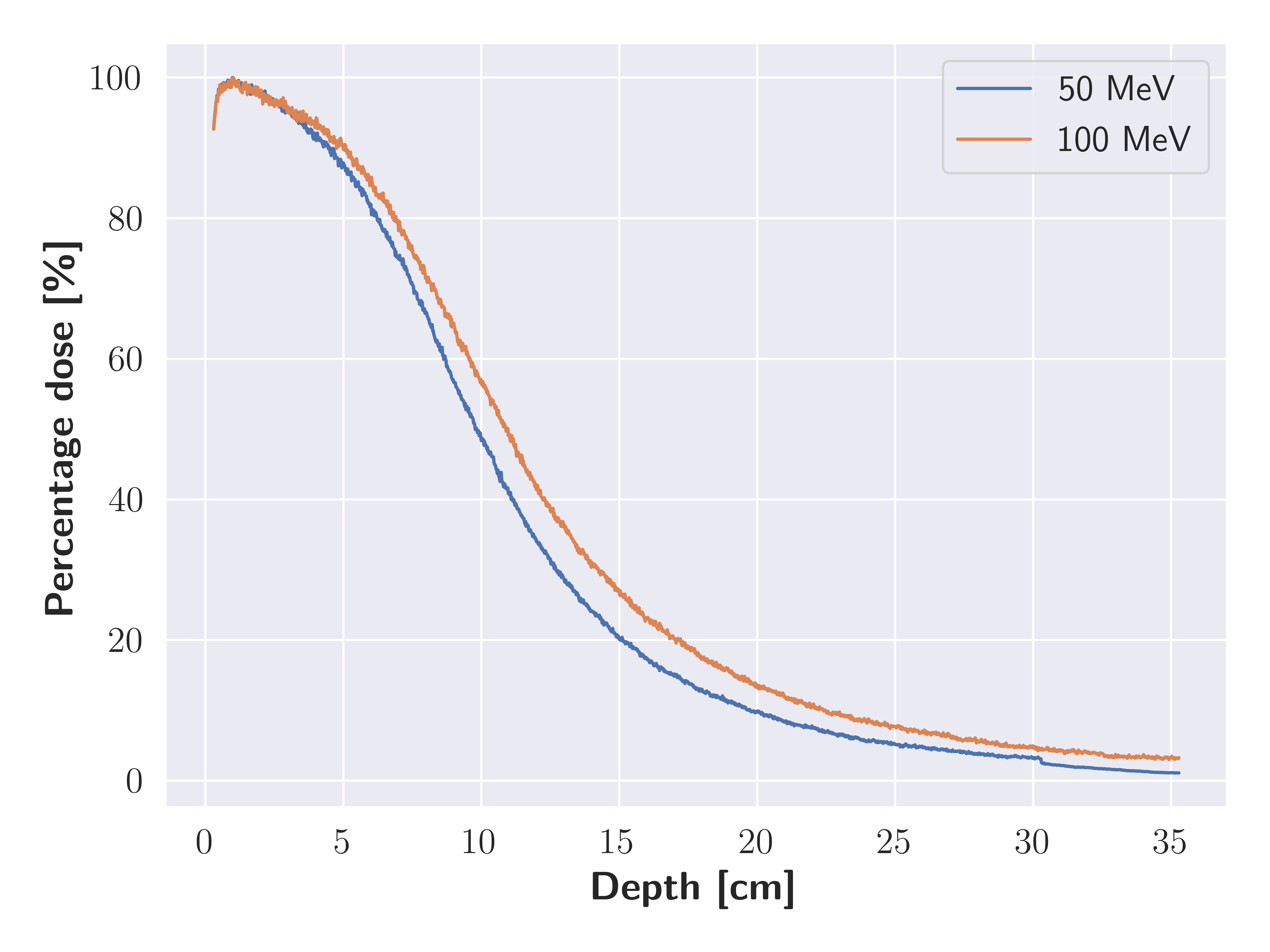}
  \end{minipage}
    \caption{\textit{left}: Sketch  of the simulated setup for beamlet dosimetry.\textit{right}: Percentage Dose Depth curves for two different values of the low-energy cut, as resulting from Monte Carlo simulations.
    }
    \label{fig::SketchAndPDD}
\end{figure}

\section*{Results}

\subsection*{Single pencil beam dosimetry}
As a first step to approach the study of a scanning procedure to cover the tumor volume, we preliminarily investigated the dosimetric features of our beam at the position of the brain model.
To this purpose, a parallelepiped water phantom (with transverse/longitudinal size of $100/300\,\mathrm{mm}$) was used.
We first observe that the fraction of the original bunch charge retained after the passage through the collimator, vacuum pipe and window corresponds to $\sim 0.15$ ($\sim 20\,\mathrm{pC}$).
In this Section, unless otherwise stated, the dose values reported refer to a single laser shot, since this is the elementary unit in the case of laser-driven accelerators.\par

Figure \ref{fig::SketchAndPDD} \textit{right} shows the on-axis Percentage Dose Depth (PDD) curve for two cases, corresponding to different values of the low energy threshold we considered to remove the "low" energy components of the beam as resulting from the PIC simulations.
As mentioned above, although the LWFA regime was optimized to yield most of the electron charge in the energy range greater than $100\,\mathrm{MeV}$, it still exhibits a non-negligible charge at lower energy (see SM for further details).
As anticipated, a post-acceleration transport and low-energy rejection beamline is therefore proposed.
Although the PDD curves do not appear to dramatically differ in the two cases, electrons with energy below $100\,\mathrm{MeV}$ actually suffer from a higher transverse scattering both in air (i.e., on exiting the vacuum line) and within the patient brain.
To minimize the effect of the transverse scattering we thus consider in the following an electron beam with a low energy cut at $100\,\mathrm{MeV}$.
This is an important consideration when planning VHEE accelerators where a trade off between compactness and beam energy is necessary, like in the case of conventional RF accelerators where the size scales linearly with the beam energy.\par

As expected for VHEE electrons 
the on-axis PDD exhibits a shallow maximum at a depth below $5\,\mathrm{mm}$, dropping quite rapidly with depth due to the transverse scattering. 
In fact, as it is already known\cite{Damiano:libroDiTesto}, a PDD dependence on the field dimension can be actually observed
when the penetration range of the electrons is greater than the field size.
In this condition the lateral particle equilibrium is lost and fewer particles deposit dose on the beam central axis, leading to an apparent shallower dose maximum when the PDD is retrieved.
Looking at the PDD curve, the $R_{80}$ (i.e., the depth at which the dose value drops at $80\%$ of the maximum), can be estimated to be $R_{80}\approx 70\,\mathrm{mm}$.
This parameter plays a peculiar role in current protocols, as it is considered the maximum range of the beam which has therapeutic relevance in conventional radiotherapy.
According to the previous considerations, however, this point deserves further investigation when dealing with mm-sized beams, as the PDD retrieved for such beams may be only partially linked to the useful depth of irradiation.

\begin{figure}[h!]
    \centering
    \begin{tabular}{cc}
    \includegraphics[width=0.45\textwidth]{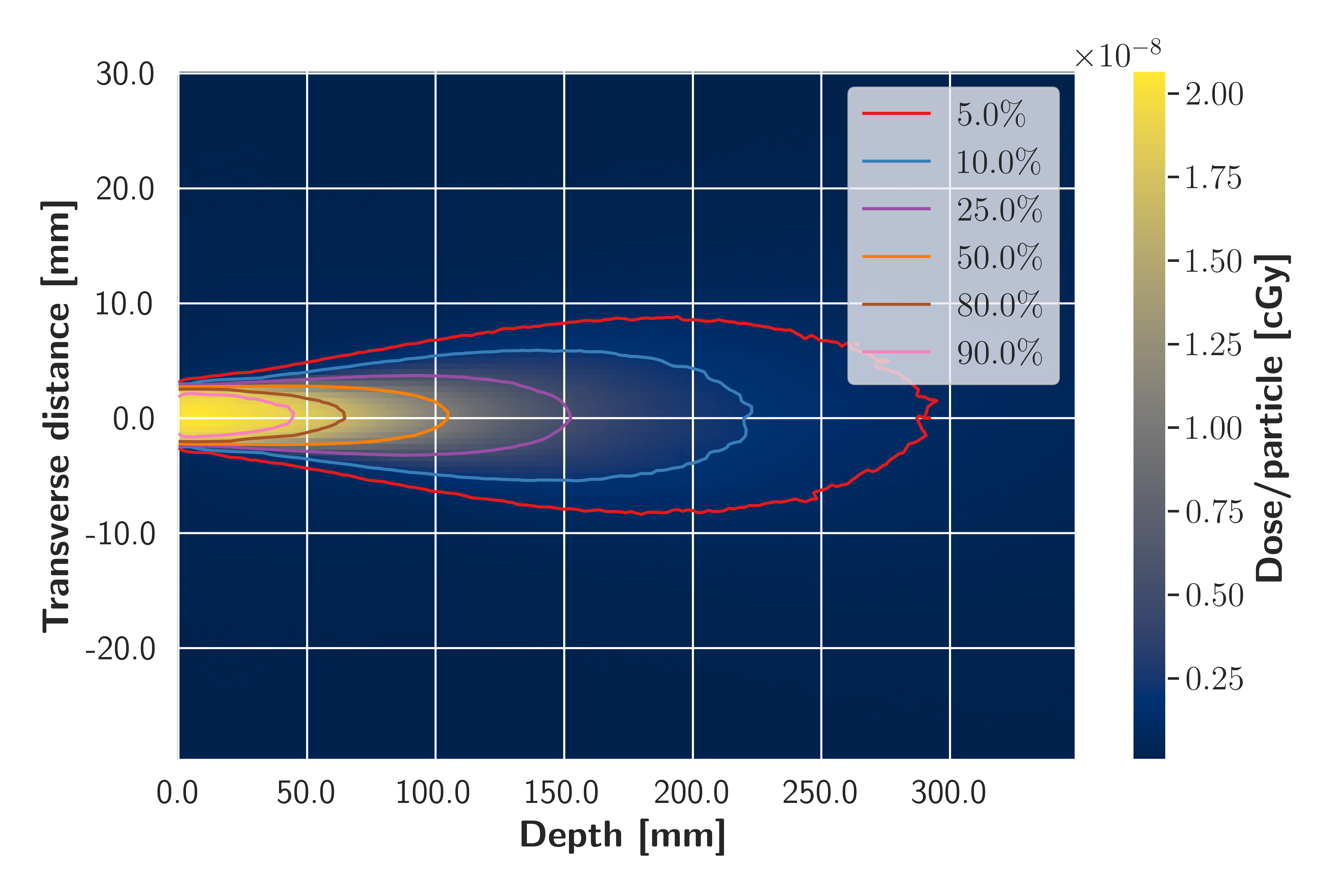}
    &
    \includegraphics[width=0.4\textwidth]{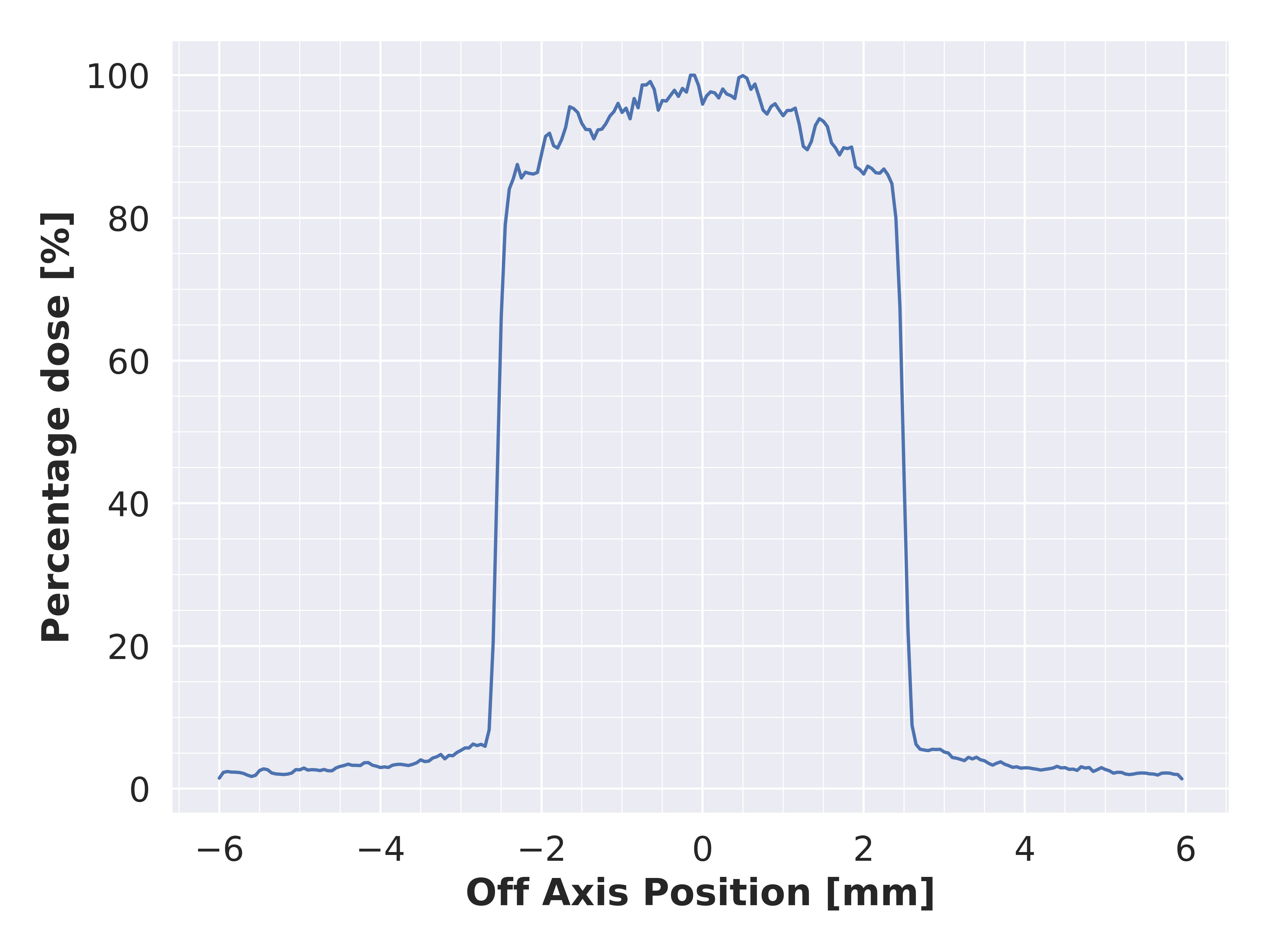}
    \\
    \end{tabular}
    \caption{
    \textit{Left}: Transverse dose distribution map, with isodose curves.
    \textit{Right}:
    Transverse beamlet profile.}
    \label{fig:DoseProfiles}
\end{figure}

Figure \ref{fig:DoseProfiles} \textit{left} shows a cut of the dose deposition map along a longitudinal plane.
As it can be seen, the transverse size of the $5\%$ isodose curve of the beamlet is confined within $10\,\mathrm{mm}$; such a relatively low transverse spreading was already observed at the beginning of the 2000s, using MC simulations, for electrons with energy above $\sim 100\,\mathrm{MeV}$\cite{Papiez_TCRT2002,Yeboah_PMB2002}, and is essential to achieve a good dose conformation for deep seated tumors.
The maximum local value of the dose is $D_{max}\simeq 2 \cdot 10^{-8}\,\mathrm{cGy/particle}$.
We refer to the final section for a deeper discussion on this point, and in particular on the possibility for laser-driven LWFA accelerators to provide UHDR, FLASH-ready, VHEE beams in the future.
Finally, Figure 3 \textit{right} shows a transverse dose profile at the position of the maximum PDD value (buildup).
Following the American Association of Medical Physics task group report 25 \cite{Khan_MP1991}, we calculated the size, symmetry and flatness of the beamlet (see Table \ref{table:BeamParameters}).
The dose profile at $z_{max}$ has a size of $5\times5\,\mathrm{mm^2}$, with flatness and symmetry deviation under $10\%$.
Although this is already a rather good level, it may seem sub-optimal with respect to current RT standards.
However, since our work is intended to establish only a first block toward more advanced dose conformation techniques with such novel pencil beams, possibly using different configurations and techniques with respect to the current protocols, we choose not to address the issue of a further optimization of these values at this stage of the work.

\subsection*{Brain tumor scanning dosimetry}
Here we present the results of the simulated irradiation of a simplified deep seated brain tumor, whose coverage was carried out with a scanning technique using the pencil beam characterized above.
The beamlets were sent to the patient along a common direction, which in the following will be referred to as the irradiation direction; in particular, we have chosen the so-called anterior-posterior axis for this study.
The tumor was modelled as a $10\,\mathrm{cc}$ sphere, placed at $80\,\mathrm{mm}$ depth in the CT reconstructed head of a male patient, toward the upper part of the brain (this is a quite common position for GBM, for instance); from now on, as it is customary, we will refer to this sphere as the Planned Target Volume (PTV).
Figure 4 shows the 3D views of our model, with the PTV highlighted in red.
Details on the CT model, the DICOM image file management, as well as on an experimental validation of the Monte Carlo code, can be found in M\&M.
\par

The scanning technique (namely, the total number and positioning of beamlets) was designed with the following two objectives, both to be accomplished on the proximal target projection plane that is, the projection plane of the PTV at the beam(s) entrance: a) let neighbor beamlets to "touch", that is to cover the entire plane with no or minimal overlapping, on this plane; b) make each point of the entire PTV cross section to be covered by at least one beamlet.
This procedure resulted in a total of $50$ beamlets needed to cover the PTV.
\par

Figure 5 shows the dose maps as resulting from this pencil beam scanning.
We first observe that our procedure leads to unwanted dose overshoots upstream of the PTV, where overlapping among the dose from neighbor beamlets occurs; in our study, the maximum dose across overlapping regions could be estimated to reach a maximum value of $\sim 119\%$ of the dose along the center of each beamlet.
This is a direct consequence of our choice of the coverage condition on the proximal plane and can be further optimized using more advanced scanning patterns.
Furthermore, the unwanted dose in the proximal region reaches a value up to $\sim 105\%$ of the maximum dose in the PTV; this is a direct effect of both the usage of a broad spectrum and of a rather small section beam; as it happens with current photon-based protocols, this unwanted effect can be greatly mitigated using advanced conformation techniques.
\par

As it can be seen, the target coverage and homogeneity are quite good ($D_{mean}\approx 98\%$, $D_{max} \approx 108\%$ considering the dose distribution normalized to cover 95\% of the target volume with 95\% of an arbitrary prescription dose).
This is a remarkable outcome, even in light of the fact that our PTV was positioned in a deep position, nearly at the center of the head.

It is also worth observing that the dose transverse spread is limited to a few millimeters from the centers of the most external beamlets; this is made possible by the usage of VHEE beams, whose well known plus in terms of reduced transverse scattering is preserved in our pencil beam scan pattern.

Finally, all the previous observations are summed up in the Dose Volume Histogram shown in Figure 6.
Beside the PTV, three more regions of interest were identified, namely the whole head, the whole brain, and an anterior-posterior section of the brain hosting the PTV, centered at the center of the PTV itself and extending beyond its edges for  $20\,\mathrm{mm}$ in the horizontal direction and for $10\,\mathrm{mm}$ in the vertical one (see Figure 5); the latter volume allows a better assessment of the lateral dose spread as well of the unwanted dose in the proximal area.
Beside a conventional radiation treatment, we also consider here the possible contribution of the FLASH effect.
As from previous literature, we introduce a FLASH "dose modifying factor" DMF; in our case, we consider a conservative value $\mathrm{DMF} = 0.80$.
The DVH was normalized in order to obtain a 95\% PTV volume coverage with 95\% of the deposited dose.
From the DVH we can deduce that thanks to the pencil beam scanning the dose was well conformed on the target.
The volume with the second highest coverage was the \textit{ad hoc} brain section volume created around the PTV.
In this volume, in particular, we obtained an amount of volume irradiated with a high amount of dose and a $D_{max}$ of the order of within 5\% of the PTV maximum dose.
It is worth noticing, however, that, as it can be seen from the corresponding curves with $\mathrm{DMF}=0.8$, taking into account a sparing due to the FLASH effect greatly improves the qualify of these DVH, even in spite of the fact that we are actually considering a highest DMF value with respect to what is emerging from recent experiments \cite{Bohlen_IJoRO2022}.

\begin{figure}[h!]
    \centering
    \includegraphics[width=0.7\textwidth]{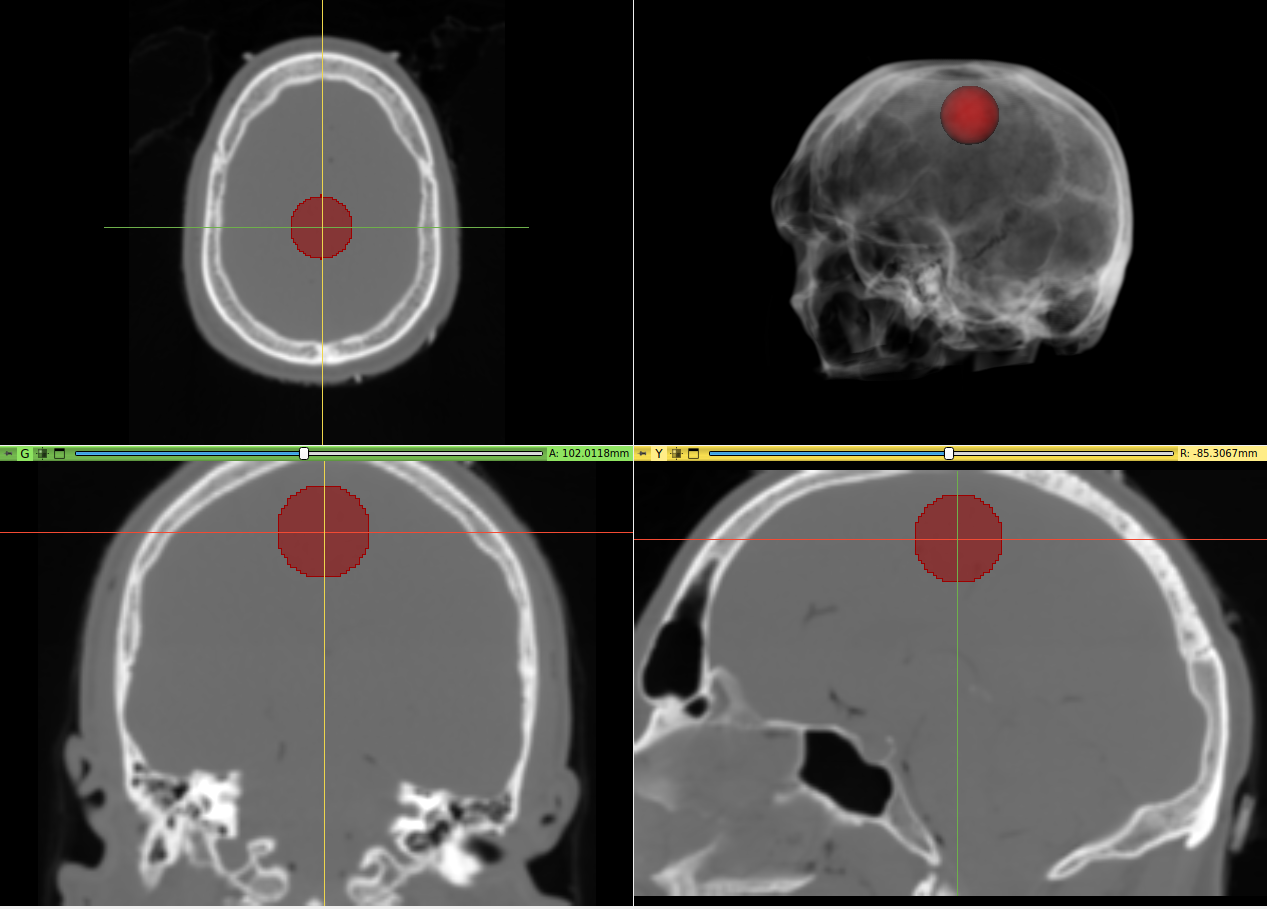}
    \caption{Ad hoc placed tumor volume (red sphere of $\sim$10 cc) in the reconstructed CT volume; projections and 3D view.}
    \label{fig:4views}
\end{figure}

\begin{figure}[h!]
    \centering
    \includegraphics[width=0.7\textwidth]{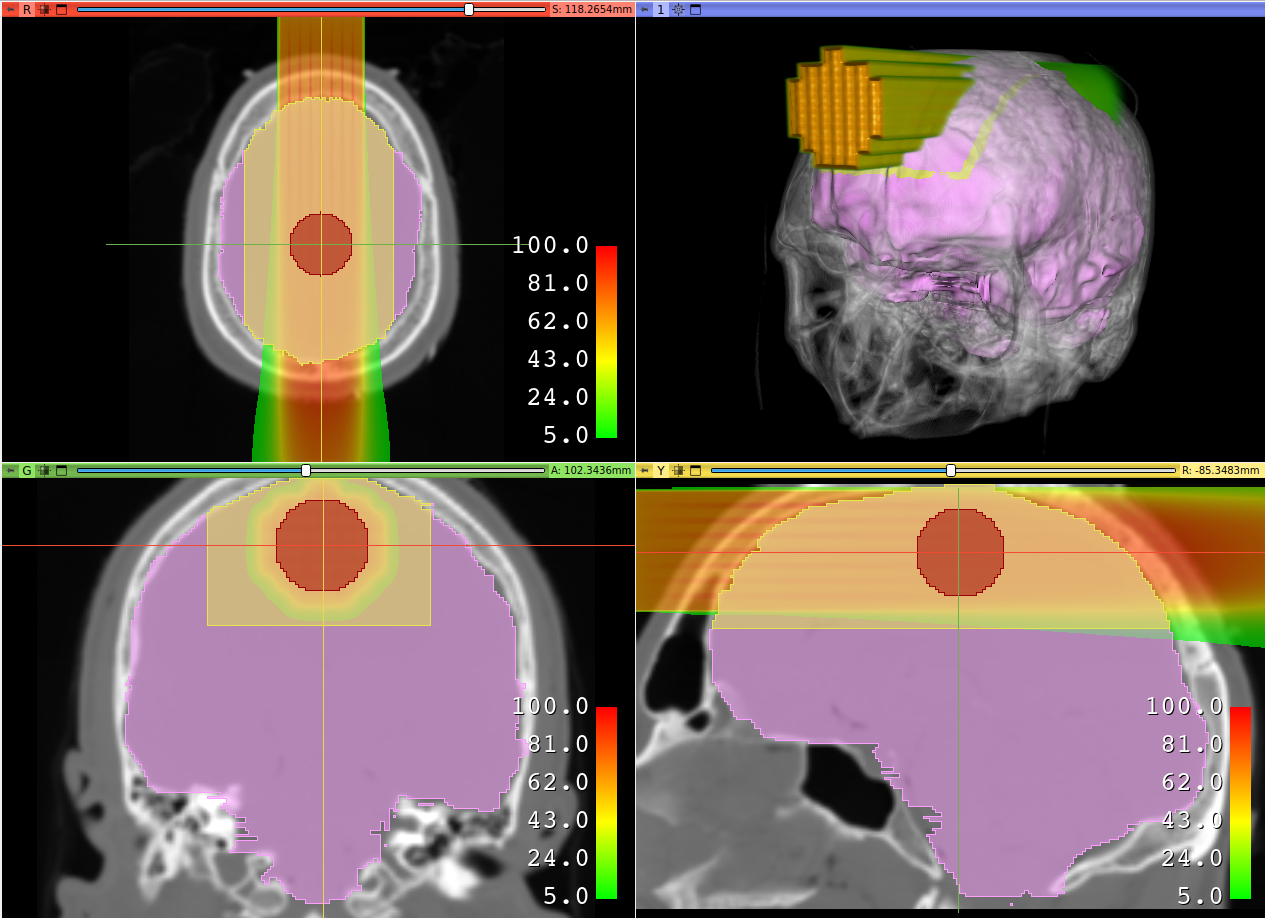}
    \caption{Dose deposition of the pencil beam scanning simulation. PTV and organs-at-risks segmentations are superimposed to the CT image.}
    \label{fig:4views_dose}
\end{figure}

\begin{figure}[h!]
    \centering
    \includegraphics[width=0.7\textwidth]{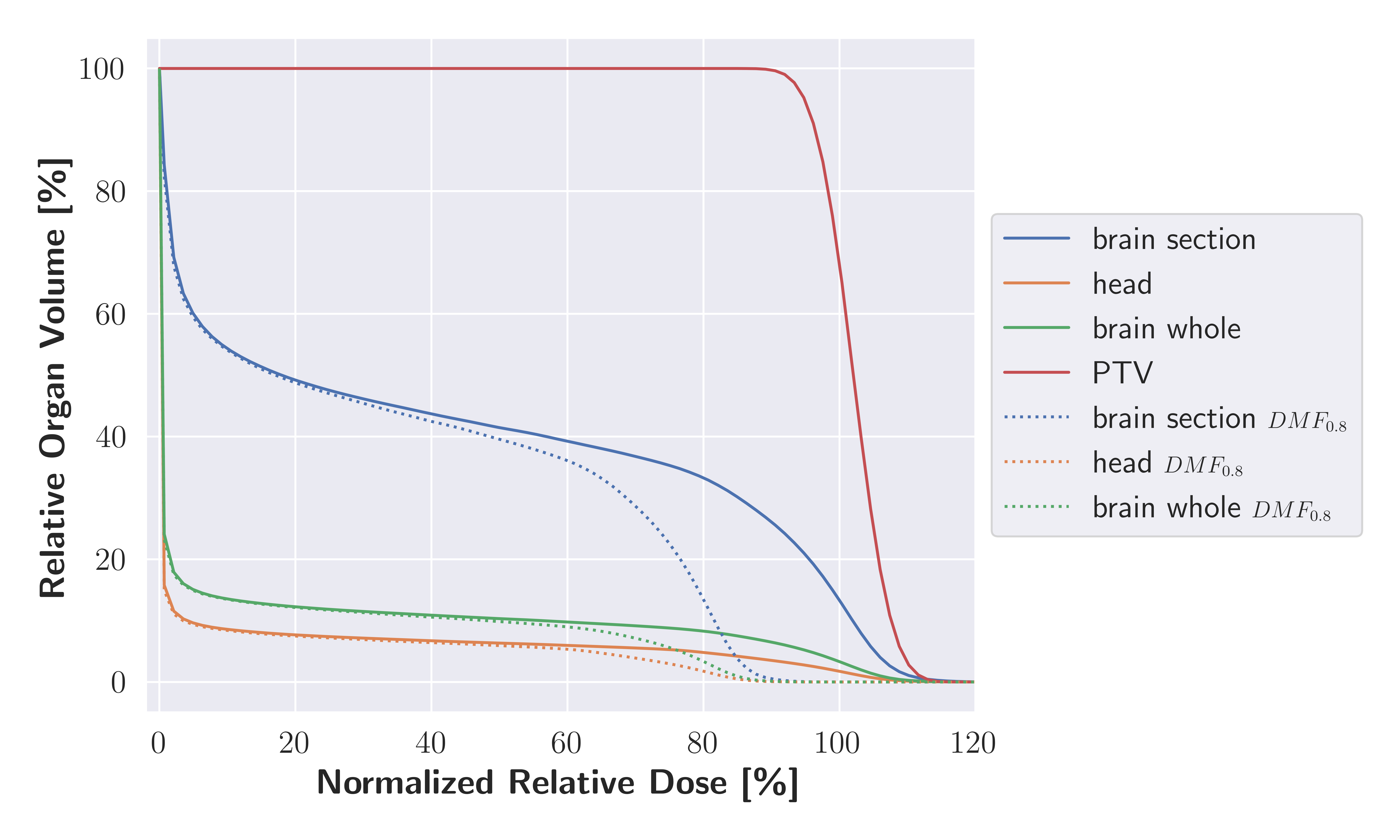}
    \caption{Dose Volume Histogram of the irradiated areas.}
    \label{fig:dvh}
\end{figure}

\section*{Perspectives for laser-driven VHEE RT and open issues}
In this work, we have investigated via Monte Carlo simulations the potential of a typical laser-driven VHEE pencil beam for the irradiation of a deep brain tumor via a scanning procedure.
Although no attempt has been made here to optimize the dose pattern via multi-field conformation techniques, our study provides a much needed building block for future, more elaborated treatments.
In particular, it is worth emphasizing that, to our knowledge, this is the first work in which the dosimetry in real phantom of beams with both relatively broad spectrum and small transverse size has been investigated numerically;
our work was mostly motivated by, and designed upon, the usage of laser-driven VHEE beams, although the beam properties explored here are most likely to be expected in future UHDR beams, due to the need to greatly increase the local dose per bunch.\par

The resulting overall performance as coming out from our study looks promising and instructive, in view of the improvements which are expected to be made possible by advanced conformation techniques.
In particular, the ballistic of the beam appears very good, providing a first indication that a quite broad spectrum can be tolerated in real treatments; from this point of view, it is worth stressing that the rather important dose contribution at the entrance is expected to be much smoothed out using, for instance, multi-field irradiation, very much as in the case of current VMAT technology.
One of the issues emerging from our study pertains to the dose overshoot in the overlapping region among neighbor beamlets; this is closely related to the usage of small transverse size beams.
With this respect, smarter approaches with respect to the one adopted here, where a relatively gross padding on the proximal plane was sought for, have to be elaborated.
This is of a particular relevance in view of the increasing role that pencil particle beams are expected to play in the context of FLASH RT \cite{Vozenin_2025_ps}.
It is worth mentioning at this point that a major difference exists between VHEE and photon beams with respect to the dose transverse spreading, whose effect on the methods to cover a PTV with adjacent pencil beams have to be better investigated.

Furthermore, as anticipated above, the usage of the PDD (and the quantities derived from that) as a figure of merit of very small size VHEE beams has to be deepened; indeed, while the loss of the transverse equilibrium (due to multiple scattering) which leads to a reduced PDD depth (in terms, for instance, of $R_{80}$ or similar) is already well known, new figures need to be identified to evaluate the ability of a beam to treat deep seated tumors, as shown by this work, where a PTV beyond the "conventional" $R_{80}$ depth was covered. \par

In addition to all these considerations, it is clear that the exploitation of the FLASH effect is a strong motivation for developing laser-driven VHEE accelerators.
As it appears from the DVHs we have shown here, taking into account healthy tissue sparing due to the FLASH effect greatly improves the irradiation outcome, even in the rather simplified irradiation geometry considered here.
With this respect we mention that more accurate consideration of the FLASH effect may be introduced in the DVH calculations with respect to the one used here, such as, for instance a tissue/organ-dependent DMF or a local dose threshold for the onset of the effect.
This is, however, a rather challenging issue at this stage, since studies of the biology underpinning the effect are ongoing and questions are still open.
On the other hand, it is worth stressing that the value of the DMF we have used above is rather conservative, in light of existing studies (see for instance \cite{Bohlen_IJoRO2022} and Refs. therein).\par

Finally, we wish to briefly discuss perspectives for real FLASH-ready VHEE sources based on LWFA to become available.
As mentioned in the introduction, laser-driven electron accelerators are considered a promising route toward UHDR beam treatments of deep seated tumors, due to their intrinsic ultra-short pulse features and other outstanding features, such as, among others, the small footprint of the accelerating structure (the plasma) due to the very high accelerating gradient, which results in a compact size, thereby ultimately requiring a much small radioprotection bunker and/or overall deployment and maintenance costs.

A major effort is ongoing to further develop and engineer LWFA accelerators for VHEE and FLASH radiotherapy.
Developments concern enhancement of single bunch properties aimed at increasing bunch charge while limiting energy spread and emittance at the levels required for proper beam transport and focusing, using dipoles, quadrupoles or sextupoles (see for instance \cite{Gaidos_2021}).
Indeed, differently from photons, VHEE beams with modest quality can be easily focused to achieve localized dose deposition, with reduced dose at the entrance and at the organs at risk surrounding the tumor.
This is somewhat similar to the case of Bragg peak with protons or light ions, but without the instrumental complexity of the hadron therapy and with much reduced sensitivity to tissue inhomogeneities in the beam path.
As it comes out from this study, with the values of charge per bunch (i.e., per laser shot) typical of the LWFA experiments carried out so far, lying in the range from a few tens up to a few hundreds of pC, the typical dose value per shot ranges from a few cGy to several tens of cGy. Recent experiments with nano-particle assisted LWFA have already demonstrated nC-level total charge \cite{aniculaesei_acceleration_2023} at even larger energies than VHEE.
On the other hand, a better energy spread control is mainly linked to the control of electron injection, which is a highly advanced field.
Studies aimed at this optimization having in mind FLASH application of LWFA are needed and increasing effort is ongoing.\par

Engineering of LWFA sources is mainly aimed at increasing the pulse repetition rate to meet the VHEE and FLASH constraint and ensure a stable and reproducible beam generation.
In fact, delivering a therapeutic dose of $\BigOSI{10} {Gy}$ for VHEE radiotherapy requires accumulation of hundreds of laser shots whose statistical properties are crucial to ensure compliance with radiotherapy standards.
Moreover, in the case of FLASH radiotherapy, delivering the above dose in the time window required for activation of the FLASH effect, $\approx$500 ms, requires a high operational repetition rate of LWFA, $\BigOSI{100} {Hz}$ or above.
So far, kHZ repetition rate has been demonstrated with MeV scale LWFA, while for VHEE beams the demonstrated repetition rate is limited to lower values, around 1 to 10 Hz.
Such limitation is mainly due to the available industrial laser systems, whose repetition rate for the pulse energy required for VHEE beams is limited to 10 Hz or less.
More recently this paradigm is changing, with the implementation of diode-pumping laser technology.
Titanium-Sapphire based commercial systems operating at Joule energy per pulse and 100 Hz repetition rate are now available specifically for the scope of VHEE radiotherapy development and other industrial applications requiring high average laser power.
At the same time, fully diode-pumped systems with new lasing materials are being demonstrated for efficient operation at the kHz repetition rate and beyond.
These performances, combined with the unique UHDR pulses make the case study of pencil beam LWFA VHEE accelerators presented here timely and highly relevant for the establishment of novel RT approaches.
Engineering developments also include gas target systems operating at a higher repetition rate or in CW, to enable LWFA operation at 100 Hz and above.
Established solutions for gas target systems already exist at an industrial level with quasi-continuous performance enabling kHz operation of LWFA accelerator, currently demonstrated at the 10 MeV electron energy \cite{PhysRevAccelBeams.23.093401}.

\section*{Materials and methods}
\subsection*{LWFA stage: PIC simulations}
The laser-plasma interaction was simulated using the Particle-In-Cell (PIC) code FBPIC\cite{LEHE201666}; this code implements a pseudo-spectral, quasi-cylindrical PIC algorithm (see SM for further details).
The driver laser, featuring a $3\,\mathrm{J}$ energy, $30\,\mathrm{fs}$ pulse duration, was focused at an intensity of $I\simeq 1.6\times 10^{19}\,\mathrm{W/cm^2}$ ($a_0\simeq 2.7$) onto a $3\,\mathrm{mm}$ long gas-jet target, made up by a mixture of He ($99\%$) and Ar ($1\%$). 
The background electron density was tailored, after a parametric study, so that the LWFA process, occurring in the so-called "bubble" regime, yielded most of the charge in the $100-250\mathrm{MeV}$ energy range; the electron density resulting from this study was  $n_e \simeq 9.4\times 10^{17}$.
In the acceleration regime selected, most of the accelerated electrons were seen to come from ionization injection of electrons  born out of the Ar $12+$ ion specie, which allowed the total charge of electrons with $E\lesssim 50 \,\mathrm{MeV}$ to be minimized.
The two dimensional ($z-r$, being $z$ and $r$ the longitudinal and transverse coordinates, respectively) simulation box considered is a moving window with size $75\times 48\,\mathrm{\mu m^2}$ and spatial resolution $0.025\times 0.08\,\mathrm{\mu m^2}$ (corresponding to a sampling of $32\times 10$ points per $\lambda$).
The field expansion in azimuthal modes was truncated to second order, and the time resolution was $\Delta t= 0.08\,\mathrm{fs}$, with a total time of $T\simeq 10\,\mathrm{ps}$.

\subsection*{VHEE beam dosimetry and tumor scanning: Monte Carlo simulations and experimental benchmarking}

The transport and dosimetric properties of the laser-driven VHEE beam obtained by the PIC simulations were studied using a homemade code based on the GEANT4 toolkit (version 10.7p3)\cite{GEANT4}.
The same code, which features multi-threading capabilities and is equipped with a DICOM file importing routine, was used to study and optimize the PTV scanning procedure in the brain model.
Transport and interactions of the particles were simulated using the QBBC physics list with the electromagnetic physics constructor 4, the most accurate for low energy particle transport; the particle range was set to $100 \; \mu\text{m}$ in all regions. 
The primary particles phase space was generated by an inversion sampling method applied to the energy and angular distributions obtained from the PIC simulations. 
A total of up to $10^7$ primary particles were typically used for each run.\par

The study of the PTV scanning technique with our laser-driven pencil beams was performed using a real head region CT of a male patient taken from the Visible Human Project database\cite{VHPwebsite}.
The DICOM CT file hosted an axial CT reconstructed head, which was imported in the code as a set of cubic, $1\,\mathrm{mm^3}$ size voxels. 
Code-wise, this was achieved using a concrete C++ instance of the \verb+G4VNestedParameterisation+ virtual class, with noticeable memory saving with respect to a standard parameterisation.
The scored quantity was the deposited dose; the dose tally was performed on voxels of $0.5\times0.5\times 0.2$ and $0.1\times0.1\times 0.1\,\mathrm{mm^3}$ for the calculation of the PDD and of the beam profile, respectively, while the entire voxel was used for the patient head.\par

In order to assess the validity of our simulation procedure (including, in particular, the DICOM file importing routine and correct volumes representation) a benchmark was carried out of its prediction with the results of an experiment carried out using a $200\,\mathrm{TW}$ ultrashort/ultraintense laser system.
In the experiment, a laser-driven VHEE beam was first used to irradiate a cylindrical PMMA phantom, using a multi-field irradiation scheme to locally enhance the local dose at the phantom center.
The same phantom was then CT scanned, and the resulting CT DICOM file was used to simulate the dose deposition pattern, in the actual experimental conditions, with our code. 
Further details can be found in the SM.\par

The CT scan of the phantom was performed using a Optima CT540 tomograph (GE Healthcare, Chicago, USA). 
Images were acquired in helical geometry, with a pitch of 0.56 and slice spacing of $0.625\,\mathrm{mm}$. 
The X-ray tube settings were $80\,\mathrm{kVp}$, $242\,\mathrm{mA}$, and the gantry rotation speed was set at $0.7\,\mathrm{s/rotation}$. 
Image reconstruction was performed with Filtered Backprojection (FBP) with standard kernel, on a matrix of $512\times512$ with pixel spacing of $0.49\,\mathrm{mm}$. 
The resulting CT scan data were then utilized in the simulation software to reconstruct the density voxel-by-voxel based on the Hounsfield Units (HU) obtained from the image.

\bibliography{sample,biblioLucaVHEEFLASH}

@article{bourhis2019treatment,
  title={Treatment of a first patient with FLASH-radiotherapy},
  author={Bourhis, Jean and Sozzi, Wendy Jeanneret and Jorge, Patrik Gon{\c{c}}alves and Gaide, Olivier and Bailat, Claude and Duclos, Fr{\'e}deric and Patin, David and Ozsahin, Mahmut and Bochud, Fran{\c{c}}ois and Germond, Jean-Fran{\c{c}}ois and others},
  journal={Radiotherapy and oncology},
  volume={139},
  pages={18--22},
  year={2019},
  publisher={Elsevier}
}

@article{mcgarrigle2024flash,
  title={The FLASH effect—an evaluation of preclinical studies of ultra-high dose rate radiotherapy},
  author={McGarrigle, Josie May and Long, Kenneth Richard and Prezado, Yolanda},
  journal={Frontiers in Oncology},
  volume={14},
  pages={1340190},
  year={2024}
}

@Article{Chen2022,
author={Chen, Haoran
and Han, Zhongyu
and Luo, Qian
and Wang, Yi
and Li, Qiju
and Zhou, Lisui
and Zuo, Houdong},
title={Radiotherapy modulates tumor cell fate decisions: a review},
journal={Radiation Oncology},
year={2022},
month={Dec},
day={01},
volume={17},
number={1},
pages={196},
issn={1748-717X},
doi={10.1186/s13014-022-02171-7},
url={https://doi.org/10.1186/s13014-022-02171-7}
}

@article{zhu2024global,
  title={Global radiotherapy demands and corresponding radiotherapy-professional workforce requirements in 2022 and predicted to 2050: a population-based study},
  author={Zhu, Hongcheng and Chua, Melvin Lee Kiang and Chitapanarux, Imjai and Kaidar-Person, Orit and Mwaba, Catherine and Alghamdi, Majed and Mignola, Andr{\'e}s Rodr{\'\i}guez and Amrogowicz, Natalia and Yazici, Gozde and Bourhaleb, Zouhour and others},
  journal={The Lancet Global Health},
  volume={12},
  number={12},
  pages={e1945--e1953},
  year={2024},
  publisher={Elsevier}
}

@article{RevModPhys.96.035002,
  title = {FLASH: New intersection of physics, chemistry, biology, and cancer medicine},
  author = {Vozenin, Marie-Catherine and Loo, Billy W. and Tantawi, Sami and Maxim, Peter G. and Spitz, Douglas R. and Bailat, Claude and Limoli, Charles L.},
  journal = {Rev. Mod. Phys.},
  volume = {96},
  issue = {3},
  pages = {035002},
  numpages = {50},
  year = {2024},
  month = {Sep},
  publisher = {American Physical Society},
  doi = {10.1103/RevModPhys.96.035002},
  url = {https://link.aps.org/doi/10.1103/RevModPhys.96.035002}
}

@article{Borghini2024_FLASH,
  author    = {Borghini, A. and Labate, L. and Piccinini, S. and Panaino, C. M. V. and Andreassi, M. G. and Gizzi, L. A.},
  title     = {FLASH Radiotherapy: Expectations, Challenges, and Current Knowledge},
  journal   = {International Journal of Molecular Sciences},
  year      = {2024},
  volume    = {25},
  number    = {5},
  pages     = {2546},
  doi       = {10.3390/ijms25052546}
}

@article{Chow2024_FLASH_Mechanisms,
  author    = {Chow, James C. L. and Ruda, Harry E.},
  title     = {Mechanisms of Action in FLASH Radiotherapy: A Comprehensive Review of Physicochemical and Biological Processes on Cancerous and Normal Cells},
  journal   = {Cells},
  year      = {2024},
  volume    = {13},
  number    = {10},
  pages     = {835},
  doi       = {10.3390/cells13100835}
}

@article{Rosini2025_FLASH_Mechanisms,
  author    = {Rosini, Giulia and Ciarrocchi, Esther and D'Orsi, Beatrice},
  title     = {Mechanisms of the FLASH Effect: Current Insights and Advances},
  journal   = {Frontiers in Cell and Developmental Biology},
  year      = {2025},
  volume    = {13},
  pages     = {1575678},
  doi       = {10.3389/fcell.2025.1575678}
}

@article{Bohlen2024VHEEFLASH,
  title   = {Very high-energy electron therapy as light-particle alternative to transmission proton FLASH therapy -- An evaluation of dosimetric performances},
  author  = {B{\"o}hlen, T. T. and Germond, J.-F. and Desorgher, L. and Veres, I. and Bratel, A. and Landstr{\"o}m, E. and Engwall, E. and Herrera, F. G. and Ozsahin, E. M. and Bourhis, J. and Bochud, F. and Moeckli, R.},
  journal = {Radiotherapy and Oncology},
  volume  = {194},
  pages   = {110177},
  year    = {2024},
  issn    = {0167-8140},
  doi     = {10.1016/j.radonc.2024.110177}
}

@article{Gesualdi2025ProtonVHEE,
  title   = {Comparison of protons and very high-energy electrons transmission pencil-beam-scanning for FLASH radiotherapy},
  author  = {Gesualdi, Flavia and Ermeneux, Louis and Lansonneur, Pierre and Sitarz, Mateusz and Loap, Pierre and Cr{\'e}hange, Gilles and Magliari, Anthony and De Marzi, Ludovic},
  journal = {Physics and Radiation Oncology (PHRO)},
  year    = {2025},
  doi     = {10.1016/j.phro.2025.100860},
  note    = {Transmission proton and VHEE PBS treatment plan comparison with dose rate quantification relevant to FLASH radiotherapy}  
}

@article{Kacem_IJRB2021,
	title = {Understanding the {FLASH} effect to unravel the potential of ultra-high dose rate irradiation},
	volume = {98},
	issn = {0955-3002, 1362-3095},
	url = {https://www.tandfonline.com/doi/full/10.1080/09553002.2021.2004328},
	doi = {10.1080/09553002.2021.2004328},
	language = {en},
	number = {3},
	urldate = {2023-11-26},
	journal = {International Journal of Radiation Biology},
	author = {Kacem, Houda and Almeida, Aymeric and Cherbuin, Nicolas and Vozenin, Marie-Catherine},
	month = mar,
	year = {2022},
	pages = {506--516},
}

@article{Friedl_MP2022,
	title = {Radiobiology of the {FLASH} effect},
	volume = {49},
	issn = {0094-2405, 2473-4209},
	url = {https://aapm.onlinelibrary.wiley.com/doi/10.1002/mp.15184},
	doi = {10.1002/mp.15184},
	language = {en},
	number = {3},
	urldate = {2023-12-05},
	journal = {Medical Physics},
	author = {Friedl, Anna A. and Prise, Kevin M. and Butterworth, Karl T. and Montay‐Gruel, Pierre and Favaudon, Vincent},
	month = mar,
	year = {2022},
	pages = {1993--2013},
}

@article{Wilson_FO2020a,
	title = {Ultra-{High} {Dose} {Rate} ({FLASH}) {Radiotherapy}: {Silver} {Bullet} or {Fool}'s {Gold}?},
	volume = {9},
	issn = {2234-943X},
	shorttitle = {Ultra-{High} {Dose} {Rate} ({FLASH}) {Radiotherapy}},
	url = {https://www.frontiersin.org/article/10.3389/fonc.2019.01563/full},
	doi = {10.3389/fonc.2019.01563},
	urldate = {2023-12-05},
	journal = {Frontiers in Oncology},
	author = {Wilson, Joseph D. and Hammond, Ester M. and Higgins, Geoff S. and Petersson, Kristoffer},
	month = jan,
	year = {2020},
	pages = {1563},
}

@article{Mann_FN2018,
	title = {Advances in {Radiotherapy} for {Glioblastoma}},
	volume = {8},
	issn = {1664-2295},
	url = {http://journal.frontiersin.org/article/10.3389/fneur.2017.00748/full},
	doi = {10.3389/fneur.2017.00748},
	urldate = {2023-12-05},
	journal = {Frontiers in Neurology},
	author = {Mann, Justin and Ramakrishna, Rohan and Magge, Rajiv and Wernicke, A. Gabriella},
	month = jan,
	year = {2018},
	pages = {748},
}

@book{Giulietti_2016,
	address = {Cham},
	series = {Biological and {Medical} {Physics}, {Biomedical} {Engineering}},
	title = {Laser-{Driven} {Particle} {Acceleration} {Towards} {Radiobiology} and {Medicine}},
	isbn = {978-3-319-31561-4 978-3-319-31563-8},
	url = {http://link.springer.com/10.1007/978-3-319-31563-8},
	urldate = {2023-12-05},
	publisher = {Springer International Publishing},
	editor = {Giulietti, Antonio},
	year = {2016},
	doi = {10.1007/978-3-319-31563-8},
}

@article{Farr_MP2022,
	title = {Ultra‐high dose rate radiation production and delivery systems intended for {FLASH}},
	volume = {49},
	issn = {0094-2405, 2473-4209},
	url = {https://aapm.onlinelibrary.wiley.com/doi/10.1002/mp.15659},
	doi = {10.1002/mp.15659},
	language = {en},
	number = {7},
	urldate = {2023-12-05},
	journal = {Medical Physics},
	author = {Farr, Jonathan and Grilj, Veljko and Malka, Victor and Sudharsan, Srinivasan and Schippers, Marco},
	month = jul,
	year = {2022},
	pages = {4875--4911},
}

@article{Esarey_RMP2009,
	title = {Physics of laser-driven plasma-based electron accelerators},
	volume = {81},
	issn = {0034-6861, 1539-0756},
	url = {https://link.aps.org/doi/10.1103/RevModPhys.81.1229},
	doi = {10.1103/RevModPhys.81.1229},
	language = {en},
	number = {3},
	urldate = {2023-12-05},
	journal = {Reviews of Modern Physics},
	author = {Esarey, E. and Schroeder, C. B. and Leemans, W. P.},
	month = aug,
	year = {2009},
	pages = {1229--1285},
}

@article{Esplen_PMB2020,
	title = {Physics and biology of ultrahigh dose-rate ({FLASH}) radiotherapy: a topical review},
	volume = {65},
	issn = {0031-9155, 1361-6560},
	shorttitle = {Physics and biology of ultrahigh dose-rate ({FLASH}) radiotherapy},
	url = {https://iopscience.iop.org/article/10.1088/1361-6560/abaa28},
	doi = {10.1088/1361-6560/abaa28},
	number = {23},
	urldate = {2023-12-07},
	journal = {Physics in Medicine \& Biology},
	author = {Esplen, Nolan and Mendonca, Marc S and Bazalova-Carter, Magdalena},
	month = dec,
	year = {2020},
	pages = {23TR03},
}

@article{Sampayan_SR2021,
	title = {Megavolt bremsstrahlung measurements from linear induction accelerators demonstrate possible use as a {FLASH} radiotherapy source to reduce acute toxicity},
	volume = {11},
	issn = {2045-2322},
	url = {https://www.nature.com/articles/s41598-021-95807-9},
	doi = {10.1038/s41598-021-95807-9},
	language = {en},
	number = {1},
	urldate = {2023-12-07},
	journal = {Scientific Reports},
	author = {Sampayan, Stephen E. and Sampayan, Kristin C. and Caporaso, George J. and Chen, Yu-Jiuan and Falabella, Steve and Hawkins, Steven A. and Hearn, Jason and Watson, James A. and Zentler, Jan-Mark},
	month = aug,
	year = {2021},
	pages = {17104},
}

@article{Lempart_RaO2019,
	title = {Modifying a clinical linear accelerator for delivery of ultra-high dose rate irradiation},
	volume = {139},
	issn = {01678140},
	url = {https://linkinghub.elsevier.com/retrieve/pii/S0167814019300593},
	doi = {10.1016/j.radonc.2019.01.031},
	language = {en},
	urldate = {2023-12-07},
	journal = {Radiotherapy and Oncology},
	author = {Lempart, Michael and Blad, Börje and Adrian, Gabriel and Bäck, Sven and Knöös, Tommy and Ceberg, Crister and Petersson, Kristoffer},
	month = oct,
	year = {2019},
	pages = {40--45},
}

@article{Rahman_IJoRO2021,
	title = {Electron {FLASH} {Delivery} at {Treatment} {Room} {Isocenter} for {Efficient} {Reversible} {Conversion} of a {Clinical} {LINAC}},
	volume = {110},
	issn = {03603016},
	url = {https://linkinghub.elsevier.com/retrieve/pii/S0360301621000249},
	doi = {10.1016/j.ijrobp.2021.01.011},
	language = {en},
	number = {3},
	urldate = {2023-12-07},
	journal = {International Journal of Radiation Oncology*Biology*Physics},
	author = {Rahman, Mahbubur and Ashraf, M. Ramish and Zhang, Rongxiao and Bruza, Petr and Dexter, Chad A. and Thompson, Lawrence and Cao, Xu and Williams, Benjamin B. and Hoopes, P. Jack and Pogue, Brian W. and Gladstone, David J.},
	month = jul,
	year = {2021},
	pages = {872--882},
}

@article{Felici_FP2020,
	title = {Transforming an {IORT} {Linac} {Into} a {FLASH} {Research} {Machine}: {Procedure} and {Dosimetric} {Characterization}},
	volume = {8},
	issn = {2296-424X},
	shorttitle = {Transforming an {IORT} {Linac} {Into} a {FLASH} {Research} {Machine}},
	url = {https://www.frontiersin.org/article/10.3389/fphy.2020.00374/full},
	doi = {10.3389/fphy.2020.00374},
	urldate = {2023-12-07},
	journal = {Frontiers in Physics},
	author = {Felici, Giuseppe and Barca, Patrizio and Barone, Salvatore and Bortoli, Eleonora and Borgheresi, Rita and De Stefano, Silvia and Di Francesco, Massimo and Grasso, Luigi and Linsalata, Stefania and Marfisi, Daniela and Pacitti, Matteo and Di Martino, Fabio},
	month = sep,
	year = {2020},
	pages = {374},
}

@article{Ronga_Cancers2021,
	title = {Back to the {Future}: {Very} {High}-{Energy} {Electrons} ({VHEEs}) and {Their} {Potential} {Application} in {Radiation} {Therapy}},
	volume = {13},
	issn = {2072-6694},
	shorttitle = {Back to the {Future}},
	url = {https://www.mdpi.com/2072-6694/13/19/4942},
	doi = {10.3390/cancers13194942},
	language = {en},
	number = {19},
	urldate = {2023-12-07},
	journal = {Cancers},
	author = {Ronga, Maria Grazia and Cavallone, Marco and Patriarca, Annalisa and Leite, Amelia Maia and Loap, Pierre and Favaudon, Vincent and Créhange, Gilles and De Marzi, Ludovic},
	month = sep,
	year = {2021},
	pages = {4942},
}

@article{Righi_JACMP2013,
	title = {Dosimetric characteristics of electron beams produced by two mobile accelerators, {Novac7} and {Liac}, for intraoperative radiation therapy through {Monte} {Carlo} simulation},
	volume = {14},
	issn = {1526-9914, 1526-9914},
	url = {https://aapm.onlinelibrary.wiley.com/doi/10.1120/jacmp.v14i1.3678},
	doi = {10.1120/jacmp.v14i1.3678},
	language = {en},
	number = {1},
	urldate = {2023-12-07},
	journal = {Journal of Applied Clinical Medical Physics},
	author = {Righi, Sergio and Karaj, Evis and Felici, Giuseppe and Di Martino, Fabio},
	month = jan,
	year = {2013},
	pages = {6--18},
}

@article{DesRosiers_PMB2000,
	title = {150-250 {MeV} electron beams in radiation therapy},
	volume = {45},
	issn = {0031-9155, 1361-6560},
	url = {https://iopscience.iop.org/article/10.1088/0031-9155/45/7/306},
	doi = {10.1088/0031-9155/45/7/306},
	number = {7},
	urldate = {2023-12-07},
	journal = {Physics in Medicine and Biology},
	author = {DesRosiers, C and Moskvin, V and Bielajew, A F and Papiez, L},
	month = jul,
	year = {2000},
	pages = {1781--1805},
}

@article{Yeboah_PMB2002,
	title = {Optimization of intensity-modulated very high energy (50–250 {MeV}) electron therapy},
	volume = {47},
	issn = {0031-9155, 1361-6560},
	url = {https://iopscience.iop.org/article/10.1088/0031-9155/47/8/305},
	doi = {10.1088/0031-9155/47/8/305},
	number = {8},
	urldate = {2023-12-07},
	journal = {Physics in Medicine and Biology},
	author = {Yeboah, C and Sandison, G A and Moskvin, V},
	month = apr,
	year = {2002},
	pages = {1285--1301},
}

@article{Palma_RaO2016,
	title = {Assessment of the quality of very high-energy electron radiotherapy planning},
	volume = {119},
	issn = {01678140},
	url = {https://linkinghub.elsevier.com/retrieve/pii/S0167814016000487},
	doi = {10.1016/j.radonc.2016.01.017},
	language = {en},
	number = {1},
	urldate = {2023-12-07},
	journal = {Radiotherapy and Oncology},
	author = {Palma, Bianey and Bazalova-Carter, Magdalena and Hårdemark, Björn and Hynning, Elin and Qu, Bradley and Loo, Billy W. and Maxim, Peter G.},
	month = apr,
	year = {2016},
	pages = {154--158},
}

@article{BazalovaCarter_MP2015,
	title = {Treatment planning for radiotherapy with very high‐energy electron beams and comparison of {VHEE} and {VMAT} plans},
	volume = {42},
	issn = {0094-2405, 2473-4209},
	url = {https://aapm.onlinelibrary.wiley.com/doi/10.1118/1.4918923},
	doi = {10.1118/1.4918923},
	language = {en},
	number = {5},
	urldate = {2023-12-07},
	journal = {Medical Physics},
	author = {Bazalova‐Carter, Magdalena and Qu, Bradley and Palma, Bianey and Hårdemark, Björn and Hynning, Elin and Jensen, Christopher and Maxim, Peter G. and Loo, Billy W.},
	month = may,
	year = {2015},
	pages = {2615--2625},
}

@article{Fuchs_PMB2009,
	title = {Treatment planning for laser-accelerated very-high energy electrons},
	volume = {54},
	issn = {0031-9155, 1361-6560},
	url = {https://iopscience.iop.org/article/10.1088/0031-9155/54/11/003},
	doi = {10.1088/0031-9155/54/11/003},
	number = {11},
	urldate = {2023-12-07},
	journal = {Physics in Medicine and Biology},
	author = {Fuchs, T and Szymanowski, H and Oelfke, U and Glinec, Y and Rechatin, C and Faure, J and Malka, V},
	month = jun,
	year = {2009},
	pages = {3315--3328},
}

@article{Rahman_RaO2022,
	title = {{FLASH} radiotherapy treatment planning and models for electron beams},
	volume = {175},
	issn = {01678140},
	url = {https://linkinghub.elsevier.com/retrieve/pii/S016781402204230X},
	doi = {10.1016/j.radonc.2022.08.009},
	language = {en},
	urldate = {2023-12-07},
	journal = {Radiotherapy and Oncology},
	author = {Rahman, Mahbubur and Trigilio, Antonio and Franciosini, Gaia and Moeckli, Raphaël and Zhang, Rongxiao and Böhlen, Till Tobias},
	month = oct,
	year = {2022},
	pages = {210--221},
}

@article{Lagzda_NIMB2020,
	title = {Influence of heterogeneous media on {Very} {High} {Energy} {Electron} ({VHEE}) dose penetration and a {Monte} {Carlo}-based comparison with existing radiotherapy modalities},
	volume = {482},
	issn = {0168583X},
	url = {https://linkinghub.elsevier.com/retrieve/pii/S0168583X20304043},
	doi = {10.1016/j.nimb.2020.09.008},
	language = {en},
	urldate = {2023-12-07},
	journal = {Nuclear Instruments and Methods in Physics Research Section B: Beam Interactions with Materials and Atoms},
	author = {Lagzda, Agnese and Angal-Kalinin, Deepa and Jones, James and Aitkenhead, Adam and Kirkby, Karen J. and MacKay, Ranald and Van Herk, Marcel and Farabolini, Wilfrid and Zeeshan, Sumaira and Jones, Roger M.},
	month = nov,
	year = {2020},
	pages = {70--81},
}

@inproceedings{DesRosiers_SPIE2008,
	address = {San Jose, CA},
	title = {Laser-plasma generated very high energy electrons in radiation therapy of the prostate},
	url = {http://proceedings.spiedigitallibrary.org/proceeding.aspx?doi=10.1117/12.761663},
	doi = {10.1117/12.761663},
	urldate = {2023-12-07},
	author = {DesRosiers, Colleen and Moskvin, Vadim and Cao, Minsong and Joshi, Chandrashekhar J. and Langer, Mark},
	editor = {Neev, Joseph and Nolte, Stefan and Heisterkamp, Alexander and Schaffer, Christopher B.},
	month = feb,
	year = {2008},
	pages = {688109},
}

@article{Maxim_RaO2019,
	title = {{PHASER}: {A} platform for clinical translation of {FLASH} cancer radiotherapy},
	volume = {139},
	issn = {01678140},
	shorttitle = {{PHASER}},
	url = {https://linkinghub.elsevier.com/retrieve/pii/S0167814019304049},
	doi = {10.1016/j.radonc.2019.05.005},
	language = {en},
	urldate = {2023-12-07},
	journal = {Radiotherapy and Oncology},
	author = {Maxim, Peter G. and Tantawi, Sami G. and Loo, Billy W.},
	month = oct,
	year = {2019},
	pages = {28--33},
}

@article{Kokurewicz_SR2019,
	title = {Focused very high-energy electron beams as a novel radiotherapy modality for producing high-dose volumetric elements},
	volume = {9},
	issn = {2045-2322},
	url = {https://www.nature.com/articles/s41598-019-46630-w},
	doi = {10.1038/s41598-019-46630-w},
	language = {en},
	number = {1},
	urldate = {2023-12-07},
	journal = {Scientific Reports},
	author = {Kokurewicz, K. and Brunetti, E. and Welsh, G. H. and Wiggins, S. M. and Boyd, M. and Sorensen, A. and Chalmers, A. J. and Schettino, G. and Subiel, A. and DesRosiers, C. and Jaroszynski, D. A.},
	month = jul,
	year = {2019},
	pages = {10837},
}

@article{Gonsalves_PRL2019,
	title = {Petawatt {Laser} {Guiding} and {Electron} {Beam} {Acceleration} to 8 {GeV} in a {Laser}-{Heated} {Capillary} {Discharge} {Waveguide}},
	volume = {122},
	issn = {0031-9007, 1079-7114},
	url = {https://link.aps.org/doi/10.1103/PhysRevLett.122.084801},
	doi = {10.1103/PhysRevLett.122.084801},
	language = {en},
	number = {8},
	urldate = {2023-12-07},
	journal = {Physical Review Letters},
	author = {Gonsalves, A. J. and Nakamura, K. and Daniels, J. and Benedetti, C. and Pieronek, C. and De Raadt, T. C. H. and Steinke, S. and Bin, J. H. and Bulanov, S. S. and Van Tilborg, J. and Geddes, C. G. R. and Schroeder, C. B. and Tóth, Cs. and Esarey, E. and Swanson, K. and Fan-Chiang, L. and Bagdasarov, G. and Bobrova, N. and Gasilov, V. and Korn, G. and Sasorov, P. and Leemans, W. P.},
	month = feb,
	year = {2019},
	pages = {084801},
}

@article{Danson_HPLSE2019,
	title = {Petawatt and exawatt class lasers worldwide},
	volume = {7},
	issn = {2095-4719, 2052-3289},
	url = {https://www.cambridge.org/core/product/identifier/S2095471919000367/type/journal_article},
	doi = {10.1017/hpl.2019.36},
	language = {en},
	urldate = {2023-12-07},
	journal = {High Power Laser Science and Engineering},
	author = {Danson, Colin N. and Haefner, Constantin and Bromage, Jake and Butcher, Thomas and Chanteloup, Jean-Christophe F. and Chowdhury, Enam A. and Galvanauskas, Almantas and Gizzi, Leonida A. and Hein, Joachim and Hillier, David I. and Hopps, Nicholas W. and Kato, Yoshiaki and Khazanov, Efim A. and Kodama, Ryosuke and Korn, Georg and Li, Ruxin and Li, Yutong and Limpert, Jens and Ma, Jingui and Nam, Chang Hee and Neely, David and Papadopoulos, Dimitrios and Penman, Rory R. and Qian, Liejia and Rocca, Jorge J. and Shaykin, Andrey A. and Siders, Craig W. and Spindloe, Christopher and Szatmári, Sándor and Trines, Raoul M. G. M. and Zhu, Jianqiang and Zhu, Ping and Zuegel, Jonathan D.},
	year = {2019},
	pages = {e54},
}

@article{Gizzi_HPLSE2021,
	title = {Overview and specifications of laser and target areas at the {Intense} {Laser} {Irradiation} {Laboratory}},
	volume = {9},
	issn = {2095-4719, 2052-3289},
	url = {https://www.cambridge.org/core/product/identifier/S209547192000047X/type/journal_article},
	doi = {10.1017/hpl.2020.47},
	language = {en},
	urldate = {2023-12-07},
	journal = {High Power Laser Science and Engineering},
	author = {Gizzi, Leonida A. and Labate, Luca and Baffigi, Federica and Brandi, Fernando and Bussolino, Giancarlo and Fulgentini, Lorenzo and Köster, Petra and Palla, Daniele},
	year = {2021},
	pages = {e10},
}

@article{Maier_PRX2020,
	title = {Decoding {Sources} of {Energy} {Variability} in a {Laser}-{Plasma} {Accelerator}},
	volume = {10},
	issn = {2160-3308},
	url = {https://link.aps.org/doi/10.1103/PhysRevX.10.031039},
	doi = {10.1103/PhysRevX.10.031039},
	language = {en},
	number = {3},
	urldate = {2023-12-07},
	journal = {Physical Review X},
	author = {Maier, Andreas R. and Delbos, Niels M. and Eichner, Timo and Hübner, Lars and Jalas, Sören and Jeppe, Laurids and Jolly, Spencer W. and Kirchen, Manuel and Leroux, Vincent and Messner, Philipp and Schnepp, Matthias and Trunk, Maximilian and Walker, Paul A. and Werle, Christian and Winkler, Paul},
	month = aug,
	year = {2020},
	pages = {031039},
}

@article{Labate_SR2020,
	title = {Toward an effective use of laser-driven very high energy electrons for radiotherapy: {Feasibility} assessment of multi-field and intensity modulation irradiation schemes},
	volume = {10},
	issn = {2045-2322},
	shorttitle = {Toward an effective use of laser-driven very high energy electrons for radiotherapy},
	url = {https://www.nature.com/articles/s41598-020-74256-w},
	doi = {10.1038/s41598-020-74256-w},
	language = {en},
	number = {1},
	urldate = {2023-12-07},
	journal = {Scientific Reports},
	author = {Labate, Luca and Palla, Daniele and Panetta, Daniele and Avella, Federico and Baffigi, Federica and Brandi, Fernando and Di Martino, Fabio and Fulgentini, Lorenzo and Giulietti, Antonio and Köster, Petra and Terzani, Davide and Tomassini, Paolo and Traino, Claudio and Gizzi, Leonida A.},
	month = oct,
	year = {2020},
	pages = {17307},
}

@article{Svendsen_SR2021,
	title = {A focused very high energy electron beam for fractionated stereotactic radiotherapy},
	volume = {11},
	issn = {2045-2322},
	url = {https://www.nature.com/articles/s41598-021-85451-8},
	doi = {10.1038/s41598-021-85451-8},
	language = {en},
	number = {1},
	urldate = {2023-12-07},
	journal = {Scientific Reports},
	author = {Svendsen, Kristoffer and Guénot, Diego and Svensson, Jonas Björklund and Petersson, Kristoffer and Persson, Anders and Lundh, Olle},
	month = mar,
	year = {2021},
	pages = {5844},
}

@article{Sarti_FO2021,
	title = {Deep {Seated} {Tumour} {Treatments} {With} {Electrons} of {High} {Energy} {Delivered} at {FLASH} {Rates}: {The} {Example} of {Prostate} {Cancer}},
	volume = {11},
	issn = {2234-943X},
	shorttitle = {Deep {Seated} {Tumour} {Treatments} {With} {Electrons} of {High} {Energy} {Delivered} at {FLASH} {Rates}},
	url = {https://www.frontiersin.org/articles/10.3389/fonc.2021.777852/full},
	doi = {10.3389/fonc.2021.777852},
	urldate = {2023-12-07},
	journal = {Frontiers in Oncology},
	author = {Sarti, Alessio and De Maria, Patrizia and Battistoni, Giuseppe and De Simoni, Micol and Di Felice, Cinzia and Dong, Yunsheng and Fischetti, Marta and Franciosini, Gaia and Marafini, Michela and Marampon, Francesco and Mattei, Ilaria and Mirabelli, Riccardo and Muraro, Silvia and Pacilio, Massimiliano and Palumbo, Luigi and Rocca, Loredana and Rubeca, Damiana and Schiavi, Angelo and Sciubba, Adalberto and Tombolini, Vincenzo and Toppi, Marco and Traini, Giacomo and Trigilio, Antonio and Patera, Vincenzo},
	month = dec,
	year = {2021},
	pages = {777852},
}

@article{Khan_MP1991,
	title = {Clinical electron‐beam dosimetry: {Report} of {AAPM} {Radiation} {Therapy} {Committee} {Task} {Group} {No}. 25},
	volume = {18},
	issn = {0094-2405, 2473-4209},
	shorttitle = {Clinical electron‐beam dosimetry},
	url = {https://aapm.onlinelibrary.wiley.com/doi/10.1118/1.596695},
	doi = {10.1118/1.596695},
	language = {en},
	number = {1},
	urldate = {2023-12-07},
	journal = {Medical Physics},
	author = {Khan, Faiz M. and Doppke, Karen P. and Hogstrom, Kenneth R. and Kutcher, Gerald J. and Nath, Ravinder and Prasad, Satish C. and Purdy, James A. and Rozenfeld, Martin and Werner, Barry L.},
	month = jan,
	year = {1991},
	pages = {73--109},
}

@article{Pollock_PRL2011,
	title = {Demonstration of a {Narrow} {Energy} {Spread}, \textasciitilde 0.5 {GeV} {Electron} {Beam} from a {Two}-{Stage} {Laser} {Wakefield} {Accelerator}},
	volume = {107},
	issn = {0031-9007, 1079-7114},
	url = {https://link.aps.org/doi/10.1103/PhysRevLett.107.045001},
	doi = {10.1103/PhysRevLett.107.045001},
	language = {en},
	number = {4},
	urldate = {2023-12-07},
	journal = {Physical Review Letters},
	author = {Pollock, B. B. and Clayton, C. E. and Ralph, J. E. and Albert, F. and Davidson, A. and Divol, L. and Filip, C. and Glenzer, S. H. and Herpoldt, K. and Lu, W. and Marsh, K. A. and Meinecke, J. and Mori, W. B. and Pak, A. and Rensink, T. C. and Ross, J. S. and Shaw, J. and Tynan, G. R. and Joshi, C. and Froula, D. H.},
	month = jul,
	year = {2011},
	pages = {045001},
}

@article{Buck_PRL2013,
	title = {Shock-{Front} {Injector} for {High}-{Quality} {Laser}-{Plasma} {Acceleration}},
	volume = {110},
	issn = {0031-9007, 1079-7114},
	url = {https://link.aps.org/doi/10.1103/PhysRevLett.110.185006},
	doi = {10.1103/PhysRevLett.110.185006},
	language = {en},
	number = {18},
	urldate = {2023-12-07},
	journal = {Physical Review Letters},
	author = {Buck, A. and Wenz, J. and Xu, J. and Khrennikov, K. and Schmid, K. and Heigoldt, M. and Mikhailova, J. M. and Geissler, M. and Shen, B. and Krausz, F. and Karsch, S. and Veisz, L.},
	month = may,
	year = {2013},
	pages = {185006},
}

@article{Thaury_SR2015,
	title = {Shock assisted ionization injection in laser-plasma accelerators},
	volume = {5},
	issn = {2045-2322},
	url = {https://www.nature.com/articles/srep16310},
	doi = {10.1038/srep16310},
	language = {en},
	number = {1},
	urldate = {2023-12-07},
	journal = {Scientific Reports},
	author = {Thaury, C. and Guillaume, E. and Lifschitz, A. and Ta Phuoc, K. and Hansson, M. and Grittani, G. and Gautier, J. and Goddet, J.-P. and Tafzi, A. and Lundh, O. and Malka, V.},
	month = nov,
	year = {2015},
	pages = {16310},
}

@article{Tomassini_PRAB2019,
	title = {High quality electron bunches for a multistage {GeV} accelerator with resonant multipulse ionization injection},
	volume = {22},
	issn = {2469-9888},
	url = {https://link.aps.org/doi/10.1103/PhysRevAccelBeams.22.111302},
	doi = {10.1103/PhysRevAccelBeams.22.111302},
	language = {en},
	number = {11},
	urldate = {2023-12-07},
	journal = {Physical Review Accelerators and Beams},
	author = {Tomassini, Paolo and Terzani, Davide and Labate, Luca and Toci, Guido and Chance, Antoine and Nghiem, Phu Anh Phi and Gizzi, Leonida A.},
	month = nov,
	year = {2019},
	pages = {111302},
}

@article{Gonsalves_NP2011,
	title = {Tunable laser plasma accelerator based on longitudinal density tailoring},
	volume = {7},
	issn = {1745-2473, 1745-2481},
	url = {https://www.nature.com/articles/nphys2071},
	doi = {10.1038/nphys2071},
	language = {en},
	number = {11},
	urldate = {2023-12-07},
	journal = {Nature Physics},
	author = {Gonsalves, A. J. and Nakamura, K. and Lin, C. and Panasenko, D. and Shiraishi, S. and Sokollik, T. and Benedetti, C. and Schroeder, C. B. and Geddes, C. G. R. and Van Tilborg, J. and Osterhoff, J. and Esarey, E. and Toth, C. and Leemans, W. P.},
	month = nov,
	year = {2011},
	pages = {862--866},
}

@article{Papiez_TCRT2002,
	title = {Very {High} {Energy} {Electrons} (50 – 250 {MeV}) and {Radiation} {Therapy}},
	volume = {1},
	issn = {1533-0346, 1533-0338},
	url = {http://journals.sagepub.com/doi/10.1177/153303460200100202},
	doi = {10.1177/153303460200100202},
	language = {en},
	number = {2},
	urldate = {2023-12-07},
	journal = {Technology in Cancer Research \& Treatment},
	author = {Papiez, Lech and DesRosiers, Colleen and Moskvin, Vadim},
	month = apr,
	year = {2002},
	pages = {105--110},
}

@article{Bohlen_IJoRO2022,
	title = {Normal {Tissue} {Sparing} by {FLASH} as a {Function} of {Single}-{Fraction} {Dose}: {A} {Quantitative} {Analysis}},
	volume = {114},
	issn = {03603016},
	shorttitle = {Normal {Tissue} {Sparing} by {FLASH} as a {Function} of {Single}-{Fraction} {Dose}},
	url = {https://linkinghub.elsevier.com/retrieve/pii/S0360301622005417},
	doi = {10.1016/j.ijrobp.2022.05.038},
	language = {en},
	number = {5},
	urldate = {2023-12-07},
	journal = {International Journal of Radiation Oncology*Biology*Physics},
	author = {Böhlen, Till Tobias and Germond, Jean-François and Bourhis, Jean and Vozenin, Marie-Catherine and Ozsahin, Esat Mahmut and Bochud, François and Bailat, Claude and Moeckli, Raphaël},
	month = dec,
	year = {2022},
	pages = {1032--1044},
}

@misc{THERIQ,
    title={THERIQ company website},
    howpublished={\url{https://www.theryq-alcen.com/flash-radiotherapy-products/flashdeep/}},
    note={Accessed: 02-26-2026}
}

@misc{Lumitron,
    title={Lumitron company website},
    howpublished={\url{https://www.lumitronxrays.com/}},
    note={Accessed: 02-26-2026}}

@article{Vozenin_2025_ps,
	author = {Kacem, Houda and Kunz, Louis and Korysko, Pierre and Ollivier, Jonathan and Tsoutsou, Pelagia and Martinotti, Adrien and Rieker, Vilde and Bateman, Joseph and Farabolini, Wilfrid and Baldacchino, G{\'e}rard and Loo, Billy W. , Jr. and Limoli, Charles L. and Dosanjh, Manjit and Corsini, Roberto and Vozenin, Marie-Catherine},
	date = {2025/08/01},
	doi = {10.1016/j.radonc.2025.110942},
	isbn = {0167-8140},
	journal = {Radiotherapy and Oncology},
	journal1 = {Radiotherapy and Oncology},
	month = {2026/01/04},
	publisher = {Elsevier},
	title = {Modification of the microstructure of the CERN-CLEAR-VHEE beam at the picosecond scale modifies ZFE morphogenesis but has no impact on hydrogen peroxide production},
	type = {doi: 10.1016/j.radonc.2025.110942},
	url = {https://doi.org/10.1016/j.radonc.2025.110942},
	volume = {209},
	year = {2025},
	year1 = {2025}}

@Article{Li_FP2025,
  author   = {Li, Hongyu and Zha, Hao and Lin, Xiancai and Gao, Qiang and Liu, Focheng and Shi, Jiaru and Chen, Huaibi},
  title    = {Design of a 100-MeV compact VHEE beam line in Tsinghua University},
  journal  = {Frontiers in Physics},
  year     = {2025},
  volume   = {Volume 12 - 2024},
  issn     = {2296-424X},
  doi      = {10.3389/fphy.2024.1496272},
  url      = {https://www.frontiersin.org/journals/physics/articles/10.3389/fphy.2024.1496272},
}

@Article{Guo_NC2025,
author={Guo, Zhiyuan
and Liu, Shuang
and Zhou, Bing
and Liu, Junqi
and Wang, Haiyang
and Pi, Yifei
and Wang, Xiaoyan
and Mo, Yingyi
and Guo, Bo
and Hua, Jianfei
and Wan, Yang
and Lu, Wei},
title={Preclinical tumor control with a laser-accelerated high-energy electron radiotherapy prototype},
journal={Nature Communications},
year={2025},
month={Feb},
day={23},
volume={16},
number={1},
pages={1895},
issn={2041-1723},
doi={10.1038/s41467-025-57122-z},
url={https://doi.org/10.1038/s41467-025-57122-z}
}

@article{Favaudon_2014,
	Author = {Vincent Favaudon and Laura Caplier and Virginie Monceau and Fr{\'e}d{\'e}ric Pouzoulet and Mano Sayarath and Charles Fouillade and Marie-France Poupon and Isabel Brito and Philippe Hup{\'e} and Jean Bourhis and Janet Hall and Jean-Jacques Fontaine and Marie-Catherine Vozenin},
	Doi = {10.1126/scitranslmed.3008973},
	Eprint = {https://www.science.org/doi/pdf/10.1126/scitranslmed.3008973},
	Journal = {Science Translational Medicine},
	Number = {245},
	Pages = {245ra93-245ra93},
	Title = {Ultrahigh dose-rate FLASH irradiation increases the differential response between normal and tumor tissue in mice},
	Url = {https://www.science.org/doi/abs/10.1126/scitranslmed.3008973},
	Volume = {6},
	Year = {2014}}

@article{Borghini_2022,
	Author = {Andrea Borghini and Cecilia Vecoli and Luca Labate and Daniele Panetta and Maria Grazia Andreassi and Leonida A. Gizzi},
	Doi = {10.1080/09553002.2022.2009143},
	Eprint = {https://doi.org/10.1080/09553002.2022.2009143},
	Journal = {International Journal of Radiation Biology},
	Note = {PMID: 34913413},
	Number = {2},
	Pages = {127-135},
	Publisher = {Taylor & Francis},
	Title = {FLASH ultra-high dose rates in radiotherapy: preclinical and radiobiological evidence},
	Url = {https://doi.org/10.1080/09553002.2022.2009143},
	Volume = {98},
	Year = {2022}}

@article{globcan,
author = {Sung, Hyuna and Ferlay, Jacques and Siegel, Rebecca L. and Laversanne, Mathieu and Soerjomataram, Isabelle and Jemal, Ahmedin and Bray, Freddie},
title = {Global Cancer Statistics 2020: GLOBOCAN Estimates of Incidence and Mortality Worldwide for 36 Cancers in 185 Countries},
journal = {CA: A Cancer Journal for Clinicians},
volume = {71},
number = {3},
pages = {209-249},
doi = {https://doi.org/10.3322/caac.21660},
year = {2021}
}

@article{louis2016,
  title={The 2016 World Health Organization classification of tumors of the central nervous system: a summary},
  author={Louis, David N and Perry, Arie and Reifenberger, Guido and Von Deimling, Andreas and Figarella-Branger, Dominique and Cavenee, Webster K and Ohgaki, Hiroko and Wiestler, Otmar D and Kleihues, Paul and Ellison, David W},
  journal={Acta neuropathologica},
  volume={131},
  number={6},
  pages={803--820},
  year={2016},
  publisher={Springer}
}

@article{koshy2012,
  title={Improved survival time trends for glioblastoma using the SEER 17 population-based registries},
  author={Koshy, Matthew and Villano, John L and Dolecek, Therese A and Howard, Andrew and Mahmood, Usama and Chmura, Steven J and Weichselbaum, Ralph R and McCarthy, Bridget J},
  journal={Journal of neuro-oncology},
  volume={107},
  number={1},
  pages={207--212},
  year={2012},
  publisher={Springer}
}

@article{stupp2005,
  title={Radiotherapy plus concomitant and adjuvant temozolomide for glioblastoma},
  author={Stupp, Roger and Mason, Warren P and Van Den Bent, Martin J and Weller, Michael and Fisher, Barbara and Taphoorn, Martin JB and Belanger, Karl and Brandes, Alba A and Marosi, Christine and Bogdahn, Ulrich and others},
  journal={New England journal of medicine},
  volume={352},
  number={10},
  pages={987--996},
  year={2005},
  publisher={Mass Medical Soc}
}

@article{barani2015,
  title={Radiation therapy of glioblastoma},
  author={Barani, Igor J and Larson, David A},
  journal={Current understanding and treatment of gliomas},
  pages={49--73},
  year={2015},
  publisher={Springer}
}

@article{denunzio2020,
  title={Modern radiotherapy for pediatric brain tumors},
  author={DeNunzio, Nicholas J and Yock, Torunn I},
  journal={Cancers},
  volume={12},
  number={6},
  pages={1533},
  year={2020},
  publisher={Multidisciplinary Digital Publishing Institute}
}

@article{butler2006,
  title={Managing the cognitive effects of brain tumor radiation therapy},
  author={Butler, Jerome M and Rapp, Stephen R and Shaw, Edward G},
  journal={Current treatment options in oncology},
  volume={7},
  number={6},
  pages={517--523},
  year={2006},
  publisher={Springer}
}

@article{montay2021,
  title={Hypofractionated FLASH-RT as an effective treatment against glioblastoma that reduces neurocognitive side effects in mice},
  author={Montay-Gruel, Pierre and Acharya, Munjal M and Jorge, Patrik Gon{\c{c}}alves and Petit, Beno{\^\i}t and Petridis, Ioannis G and Fuchs, Philippe and Leavitt, Ron and Petersson, Kristoffer and Gondr{\'e}, Maude and Ollivier, Jonathan and others},
  journal={Clinical Cancer Research},
  volume={27},
  number={3},
  pages={775--784},
  year={2021},
  publisher={AACR}
}

@article{montay2017,
  title={Irradiation in a flash: Unique sparing of memory in mice after whole brain irradiation with dose rates above 100 Gy/s},
  author={Montay-Gruel, Pierre and Petersson, Kristoffer and Jaccard, Maud and Boivin, Ga{\"e}l and Germond, Jean-Fran{\c{c}}ois and Petit, Benoit and Doenlen, Rapha{\"e}l and Favaudon, Vincent and Bochud, Fran{\c{c}}ois and Bailat, Claude and others},
  journal={Radiotherapy and Oncology},
  volume={124},
  number={3},
  pages={365--369},
  year={2017},
  publisher={Elsevier}
}

@article{alaghband2020,
  title={Neuroprotection of radiosensitive juvenile mice by ultra-high dose rate flash irradiation},
  author={Alaghband, Yasaman and Cheeks, Samantha N and Allen, Barrett D and Montay-Gruel, Pierre and Doan, Ngoc-Lien and Petit, Benoit and Jorge, Patrik Goncalves and Giedzinski, Erich and Acharya, Munjal M and Vozenin, Marie-Catherine and others},
  journal={Cancers},
  volume={12},
  number={6},
  pages={1671},
  year={2020},
  publisher={Multidisciplinary Digital Publishing Institute}
}

@article{vozenin2019,
  title={Biological benefits of ultra-high dose rate FLASH radiotherapy: sleeping beauty awoken},
  author={Vozenin, M-C and Hendry, Jolyon H and Limoli, CL},
  journal={Clinical oncology},
  volume={31},
  number={7},
  pages={407--415},
  year={2019},
  publisher={Elsevier}
}

@article{LEHE201666,
title = {A spectral, quasi-cylindrical and dispersion-free Particle-In-Cell algorithm},
journal = {Computer Physics Communications},
volume = {203},
pages = {66-82},
year = {2016},
issn = {0010-4655},
doi = {https://doi.org/10.1016/j.cpc.2016.02.007},
url = {https://www.sciencedirect.com/science/article/pii/S0010465516300224},
author = {Rémi Lehe and Manuel Kirchen and Igor A. Andriyash and Brendan B. Godfrey and Jean-Luc Vay}
}

@inproceedings{Gaidos_2021,
	Author = {P. Gajdos and M. Krus},
	Booktitle = {Laser Acceleration of Electrons, Protons, and Ions VI},
	Doi = {10.1117/12.2589208},
	Editor = {Stepan S. Bulanov and J{\"o}rg Schreiber and Carl B. Schroeder},
	Organization = {International Society for Optics and Photonics},
	Pages = {117790G},
	Publisher = {SPIE},
	Title = {{Ultrashort electron beam isochronous, achromatic transport line design}},
	Url = {https://doi.org/10.1117/12.2589208},
	Volume = {11779},
	Year = {2021}}

@article{aniculaesei_acceleration_2023,
	Author = {Aniculaesei, Constantin and Ha, Thanh and Yoffe, Samuel and Labun, Lance and Milton, Stephen and McCary, Edward and Spinks, Michael M. and Quevedo, Hernan J. and Labun, Ou Z. and Sain, Ritwik and Hannasch, Andrea and Zgadzaj, Rafal and Pagano, Isabella and Franco-Altamirano, Jose A. and Ringuette, Martin L. and Gaul, Erhart and Luedtke, Scott V. and Tiwari, Ganesh and Ersfeld, Bernhard and Brunetti, Enrico and Ruhl, Hartmut and Ditmire, Todd and Bruce, Sandra and Donovan, Michael E. and Downer, Michael C. and Jaroszynski, Dino A. and Hegelich, Bjorn Manuel},
	Doi = {10.1063/5.0161687},
	Issn = {2468-2047},
	Journal = {Matter and Radiation at Extremes},
	Month = nov,
	Number = {1},
	Pages = {014001},
	Title = {The acceleration of a high-charge electron bunch to 10 {GeV} in a 10-cm nanoparticle-assisted wakefield accelerator},
	Url = {https://doi.org/10.1063/5.0161687},
	Volume = {9},
	Year = {2023}}

@article{PhysRevAccelBeams.23.093401,
  title = {Demonstration of stable long-term operation of a kilohertz laser-plasma accelerator},
  author = {Rovige, L. and Huijts, J. and Andriyash, I. and Vernier, A. and Tomkus, V. and Girdauskas, V. and Raciukaitis, G. and Dudutis, J. and Stankevic, V. and Gecys, P. and Ouille, M. and Cheng, Z. and Lopez-Martens, R. and Faure, J.},
  journal = {Phys. Rev. Accel. Beams},
  volume = {23},
  issue = {9},
  pages = {093401},
  numpages = {9},
  year = {2020},
  month = {Sep},
  publisher = {American Physical Society},
  doi = {10.1103/PhysRevAccelBeams.23.093401},
  url = {https://link.aps.org/doi/10.1103/PhysRevAccelBeams.23.093401}
}

@ARTICLE{Muscato_2023,
  
AUTHOR={Muscato, A. and Arsini, L. and Battistoni, G. and Campana, L. and Carlotti, D. and De Felice, F. and De Gregorio, A. and De Simoni, M. and Di Felice, C. and Dong, Y. and Franciosini, G. and Marafini, M. and Mattei, I. and Mirabelli, R. and Muraro, S. and Pacilio, M. and Palumbo, L. and Patera, V. and Schiavi, A. and Sciubba, A. and Schwarz, M. and Sorbino, S. and Tombolini, V. and Toppi, M. and Traini, G. and Trigilio, A. and Sarti, A.},   
	 
TITLE={Treatment planning of intracranial lesions with VHEE: comparing conventional and FLASH irradiation potential with state-of-the-art photon and proton radiotherapy},      
	
JOURNAL={Frontiers in Physics},      
	
VOLUME={11},           
	
YEAR={2023},      
	  
URL={https://www.frontiersin.org/articles/10.3389/fphy.2023.1185598},       
	
DOI={10.3389/fphy.2023.1185598},      
	
ISSN={2296-424X}
}

@Article{Panaino_2024_cancers17020181,
AUTHOR = {Panaino, Costanza Maria Vittoria and Piccinini, Simona and Andreassi, Maria Grazia and Bandini, Gabriele and Borghini, Andrea and Borgia, Marzia and Di Naro, Angelo and Labate, Luca Umberto and Maggiulli, Eleonora and Portaluri, Maurizio Giovanni Agostino and Gizzi, Leonida Antonio},
TITLE = {Very High-Energy Electron Therapy Toward Clinical Implementation},
JOURNAL = {Cancers},
VOLUME = {17},
YEAR = {2025},
NUMBER = {2},
ARTICLE-NUMBER = {181},
URL = {https://www.mdpi.com/2072-6694/17/2/181},
PubMedID = {39857964},
ISSN = {2072-6694},
DOI = {10.3390/cancers17020181}
}

For data citations of datasets uploaded to e.g. \emph{figshare}, please use the \verb|howpublished| option in the bib entry to specify the platform and the link, as in the \verb|Hao:gidmaps:2014| example in the sample bibliography file.

\section*{Acknowledgements}
We acknowledge financial support from the “Tuscany Health Ecosystem (THE) Spoke 1, Advanced Radiotherapies and Diagnostics in Oncology” funded by the NextGenerationEU (PNRR), Codice progetto ECS00000017, D.D. MUR No. 1055 23 May 2022. We also acknowledge contributions from the following projects: NextGeneration EU Integrated Infrastructure Initiative in Photonic and Quantum Sciences (IPHOQS) CUP B53C22001750006, ID D2B8D520, IR0000016; Horizon 2020 Framework Programme Research and Innovation; Program EuPRAXIA Preparatory Phase (No. 101079773); EuPRAXIA Advanced Photon Sources—EuAPS (CUP I93C21000160006, IR0000030), INFN CSN5 funded project “FRIDA”; JRA ENI-CNR on Fusion energy (CUP B34I19003070007, CNR DFM.AD006.155); EU Horizon 2020 Research and Innovation Program EuPRAXIA Preparatory Phase (No. 101079773).

\section*{Author contributions statement}
L.A.G., L.L. and S.D.S. conceived the study, L.A.G., L.L. and D.D.S. and D.P. conceptualised the MC modeling, F.A., L.L., D.D.S. and D.T. carried out the simulations, L.L., D.D.S. and L.A.G. analysed the results. S.P. and G.B. carried out dosimetric validation of the VHEE beam, L.A.G. coordinated the team. All authors reviewed the manuscript. 


The corresponding author is responsible for submitting a \href{http://www.nature.com/srep/policies/index.html#competing}{competing interests statement} on behalf of all authors of the paper. This statement must be included in the submitted article file.

\end{document}